\documentclass[12pt]{article}
\usepackage{graphicx}
\makeindex

\newcommand{\ket}[1]{| #1 \rangle}

\newcommand{\beq}{\begin{equation}}
\newcommand{\eeq}{\end{equation}}

\def\vec#1{\mathbf{#1}}

\def\xi{\mathbf{x}_i}

\begin{document}

\title{Rotational excitations of polar molecules on an optical
lattice: from novel exciton physics to quantum simulation of new
lattice models}

\author{Marina Litinskaya and Roman V. Krems\\
\it\small{Department of Chemistry, University of British Columbia,
Vancouver, V6T 1Z1, Canada}}

%\small{Department of Chemistry, University of British Columbia,
%Vancouver, V6T 1Z1, Canada}
\date{}

\maketitle

\begin{abstract}
Ultracold polar molecules trapped on an optical lattice is a
many-body system that, under appropriate conditions, may support
collective excitations reminiscent of excitons in solid state
crystals. Here, we discuss the rotational excitations of molecules
on an optical lattice leading to rotational Frenkel excitons.
Apart from solid hydrogen, there is no other natural system that
exhibits rotational excitons. The rotational excitons have unique
properties that can be exploited for tuning non-linear exciton
interactions and exciton--impurity scattering by applying an
external electric field. We show that this can be used to explore
the competing role of the dynamical and kinematic exciton--exciton
interactions in excitonic energy transfer and to study quantum
localization in a dynamically tunable disordered potential. The
rotational excitons can also be used as a basis for quantum
simulation of condensed matter models that cannot be realized with
ultracold atoms. As an example, we discuss the possibility of
engineering the Holstein model with polar molecules on an optical
lattice.
\end{abstract}

\section{Introduction}

Ultracold atoms offer the possibility to study few- and many-body
quantum systems with exquisite control over microscopic
interactions. This has led to spectacular experiments
\cite{atom-review}, bridging different areas of physics. While the
research directions stimulated by experiments with ultracold atoms
are very diverse, one can identify several overarching trends that
gained momentum in recent years. The most prominent of these
trends is the effort aimed at understanding quantum phase
transitions \cite{lincoln-carr-book}, leading to the experimental
studies of Bose--Einstein condensation (BEC) \cite{BEC}, bosonic
superfluidity \cite{SF}, quantum magnetism \cite{QMagn}, many-body
spin dynamics \cite{many-body-spin-dyn}, Efimov states
\cite{Efimov}, Bardeen--Cooper--Schrieffer (BCS) superfluidity
\cite{BCS-SF} and the BEC--BCS crossover \cite{BCS-BEC}. With the
development of techniques for trapping atoms in periodic
potentials of optical lattices \cite{trap-atom-in-OL} and single
atom detection \cite{single-atom-detection}, ultracold atoms
became an ideal platform for quantum simulation of lattice models
used in solid state physics \cite{latt-model-in-sol-st-phys}. As
demonstrated by the realization of the Mott insulator --
superfluid transition with atoms trapped in a three-dimensional
optical lattice \cite{MI-SF}, the experiments with ultracold atoms
hold the promise of insight into the details of the Hubbard model,
which may help unravel the mechanism of high-$T_{\rm c}$
superconductivity \cite{high-Tc}.

The development of experimental methods for the production of
ultracold polar molecules \cite{nj-review} has widened the
possibilities for quantum simulation of condensed matter models to
a great extent. A combination of the rotational, spin and
hyperfine degrees of freedom with the long-range dipolar
interactions enabled by the dipole moment of molecules allows for
engineering a great variety of lattice models that cannot be
realized with atoms \cite{book-chapter, N-A-spin-crystal}. Of
particular interest is the possibility of creating quantum phases
with topological order \cite{topological}, which are resilient to
perturbations preserving the topology and are, therefore, ideal
for quantum computation. To this end, a major effort of current
experiments is focused on preparing a dense ensemble of ultracold
polar molecules trapped in optical lattices \cite{pmol-in-OL-exp}.

Another focus of research with quantum degenerate gases is on
emergent phenomena, such as solitons \cite{soliton}, rotons
\cite{roton}, vortices \cite{vortex}, spin waves \cite{magnon} and
polarons \cite{polaron-exp}. These experiments aim to elucidate
emergence in natural systems, and may potentially lead to the
development of novel ultra-sensitive sensors of gravity and
electromagnetic fields. While these studies cover a wide range of
collective dynamics of ultracold atoms and molecules in a single
quantum state, much less is known about the effect of internal
degrees of freedom of ultracold particles. For example, the role
of rotational transitions in the excitation spectrum of a
molecular Bose--Einstein condensate had not been addressed until
very recently \cite{whaley-pccp}. Yet, the internal degrees of
freedom can be used to explore new regimes of collective
phenomena, especially in molecular systems that provide a dense
spectrum of internal excitations \cite{carr-molecular-hubbard,
gorshkov-superfluidity}.

In this article, we consider rotational excitations of polar
molecules trapped on an optical lattice. These excitations give
rise to rotational excitons analogous to collective electronic
excitations in molecular crystals. In contrast to excitons in
natural solids, the properties of rotational excitons can be
dynamically controlled by tuning the energy level structure of the
trapped molecules, which can be used to study new regimes of
Frenkel exciton physics not accessible in natural solid state
crystals. For example, we show that non-linear interactions of
excitons can be tuned to examine the competition between the
dynamical and kinematic effects and demonstrate that rotational
excitons can be used to study quantum localization in a
dynamically tunable disordered potential. The rotational excitons
can also be used as a basis for quantum simulation of
condensed-matter models that cannot be realized with ultracold
atoms. As an example, we discuss the possibility of engineering
the Holstein model with polar molecules on an optical lattice. In
order to present these results in the context of current research,
we briefly describe the related work on quantum simulation of
many-body Hamiltonians with ultracold atoms and molecules and the
basics of Frenkel exciton physics.

\section{Quantum simulation of lattice models}

The difficulty of simulating a quantum many-body system on a
classical computer increases exponentially with the number of
quantum states. An alternative, currently at the focus of detailed
research, is quantum simulation \cite{QS-reviews}. Quantum
simulation involves the design of a controllable quantum system in
order to simulate the properties of another, more complicated
system (analog quantum simulator). This idea is generally
attributed to Feynman \cite{feynman}, although quantum simulation
is already mentioned in an earlier publication by Manin
\cite{Manin}. The key ingredient of quantum simulation is the
mapping of the Hamiltonian of the simulated system onto the
Hamiltonian of the simulator with controllable parameters. Tuning
the parameters of the controlled system can then be used to map
out the phase diagram of the simulated system. The recent
literature offers many ideas about how to build quantum simulators
based on ultracold quantum gases \cite{QS-atoms}, ultracold
trapped ions \cite{QS-ions}, single photon sources and detectors
\cite{QS-photons}, ultracold Rydberg atoms \cite{QS-Rydberg},
circuit QED \cite{QS-circuit}, quantum dots and N--V centers in
diamond \cite{QS-NV-centers}.

Ultracold atoms trapped in a periodic potential of overlapped
laser beams (optical lattices) is considered as one of the most
promising systems for quantum simulation of lattice models
\cite{trap-atom-in-OL, review-simulation-with-atoms}. The geometry
and strength of the optical lattice potentials can be controlled
by adjusting the laser intensity and the beam overlap angle, which
enables control over the translational motion of atoms in the
lattice \cite{optical-lattice}. The separation of the lattice
sites in an optical lattice is equal to half the wavelength of the
trapping laser field. Recent technological breakthroughs permit
the detection of ultracold atoms with sub-wavelength resolution,
allowing for single-site addressing \cite{single-site-detection}.

\subsection{Lattice models}
\label{s-s-lattice models}

The most prominent example of quantum simulation with ultracold
atoms is the realization of the Hubbard model \cite{experiments,
MI-SF}
\begin{equation} \hat H_H = -t
\sum\limits_{\sigma; \langle n,m \rangle} \hat c_{n,\sigma}^\dag
\hat c_{m,\sigma} +\frac{U}{2} \sum\limits_{n} \hat
n_{n,\downarrow} \hat n_{n,\uparrow} + \sum\limits_{\sigma; n}
E_\sigma \hat n_{n, \sigma},
\end{equation}
where $\sigma = \uparrow$ or $\downarrow$, $\hat c_{n,
\sigma}^\dag$ and $\hat c_{n, \sigma}$ are the creation and
annihilation operators for fermions, satisfying the
anti-commutation rule $\{\hat c_{n, \sigma}^\dag, \hat c_{m,
\sigma}\} = \delta_{nm}$, $\hat n_{n, \sigma} = \hat c_{n,
\sigma}^\dag \hat c_{n, \sigma}$ is the particle number operator,
and $E_\sigma$ is the energy of particles in state $\sigma$. The
angle brackets $\langle n,m \rangle$ indicate that the summation
is over nearest neighbors only. In this simple form, the Hubbard
Hamiltonian accounts for nearest neighbor tunnelling ($t$) and
on-site interactions for fermions in different spin states ($U$).
The fermions with the same spin experience hard-core repulsion.
This model is widely used for the studies of high-$T_{\rm c}$
superconductivity \cite{HubMod-for-highTc} and quantum magnetism
\cite{HubMod-for-magnetism}.

With bosonic atoms, it is possible to realize the Bose-Hubbard
model:
\begin{equation}\label{H_BH}
\hat H_{BH} = -t \sum\limits_{\langle n,m \rangle} \hat b_{n}^\dag
\hat b_{m} + \frac{U}{2} \sum\limits_{n} \hat n_{n}( \hat n_{n} -1
) + E_0 \sum\limits_{n}  \hat n_{n},
\end{equation}
where $\hat b_n^\dag$ and $\hat b_n$ are the creation and
annihilation operators for bosons, satisfying the commutation rule
$[\hat b_n^\dag, \hat b_m] = \delta_{nm}$, and $\hat n_{n} = \hat
b_{n}^\dag \hat b_{n}$ is the particle number operator. This model
has been used in many studies of the properties of bosonic gases
such as superfluidity \cite{superfluidity}. If $U \ll t$, the
particles are delocalized around the lattice and the ground state
of the system is a superfluid. This Hamiltonian also allows for
modelling the properties of fermionic systems when quantum
statistics plays no role and the properties of the system are
entirely determined by the relative efficiency of hopping
(governed by $t$) and the interaction ($U$). In particular, when
$U \gg t$ (the on-site repulsion dominates over the inter-site
hopping), this bosonic Hamiltonian reproduces transition to the
Mott insulator phase \cite{mott}, where the number of atoms per
lattice site is fixed. This transition is typical for electrons in
a metal. It is thus possible to realize the same physical
phenomenon with particles satisfying different quantum statistics.
We shall discuss this in detail in section \ref{s-s-mappings}.

Recent experiments demonstrated the possibility of creating the
Mott insulator phase with ultracold atoms filling up to 99\% of
the lattice sites \cite{99-percent}. A major thrust of current
research is to extend these experiments to ultracold molecules.
The experiments with polar molecules in optical lattices offer the
possibility of realizing lattice models with long-range
interactions \cite{book-chapter, N-A-spin-crystal,
pmol-for-long-range, pmol-review}. In the Bose-Hubbard model the
inter-site interactions are accounted for by adding to the
Hamiltonian (\ref{H_BH}) a term $\hat H_V = \sum\limits_{\langle
n,m \rangle} V_{n,m} \hat n_{n} \hat n_{m}$ describing
density--density correlations. In the limit of the strong on-site
repulsion, when $\langle \hat n_n \rangle = 0$ or 1, the system
reduces to the $t$-$V$ model, described by the Hamiltonian
\begin{equation}\label{H t-V}
\hat H_{t-V} = -t \sum\limits_{\langle n,m \rangle} \hat
b_{n}^\dag \hat b_{m} + E_0 \sum\limits_{n}  \hat n_{n} + V
\sum\limits_{\langle n,m \rangle} \hat n_{n} \hat n_{m}
\end{equation}
with the constraint
\begin{equation}\label{H t-V constraint}
\hat{b}_n^\dag \hat{b}_n^\dag | \Phi \rangle = \hat{b}_n \hat{b}_n
| \Phi \rangle = 0
\end{equation}
for any physical state of the crystal $| \Phi \rangle$, which
accounts for the infinite on-site repulsion. With $V=0$ this
Hamiltonian corresponds to the lattice analog of the
Tonks-Girardeau gas \cite{tonks, girardeau}. This model is
equivalent to an anisotropic spin-$1/2$ {\it XXZ} model
\cite{1D-boson-review}.

In addition, molecules allow for the possibility of realizing a
variety of lattice spin models \cite{latt-model-in-sol-st-phys},
such as
\begin{equation}\label{H heisenberg}
\hat H_{\sigma} = -\sum\limits_{\langle n,m \rangle}(J_{x}
\sigma_n^x \sigma_m^x + J_y \sigma_n^y \sigma_m^y + J_z \sigma_n^z
\sigma_m^z) - g H\sum\limits_n {\sigma}_n^z,
\end{equation}
where $\sigma_n^{x,y,z}$ are the Pauli matrices for lattice site
$n$. This is the so-called Heisenberg model. If the interactions
are isotropic in the plane perpendicular to the $z$ axis, i.e. for
$J_x = J_y$, this Hamiltonian reduces to the {\it XXZ}-model
\cite{xxz}. In the isotropic Heisenberg model, $J_x = J_y = J_z$.
The {\it XY}-model is for the case when $J_z = 0$. The scalar
version of the Heisenberg model ($J_x = J_y = 0$) is the Ising
model. In the Ising model, spins are treated as scalars, which can
take one of two values: $s_n = \pm 1$. The Hamiltonian becomes
\begin{equation}\label{H ising}
\hat H_{I} = -\sum\limits_{n,m}J_{nm} s_n s_m - h \sum\limits_n
s_n.
\end{equation}

The Ising model can be realized with ultracold molecules in the
spin-less $^1\Sigma$ electronic state trapped on an optical
lattice in a Mott insulator phase with one molecule per lattice
site. The rotational levels of $^1\Sigma$ molecules in an external
electric field form an isolated two-level system, illustrated in
Figure \ref{f-exciton dispersion}a. The ground $| g \rangle$ and
excited $|e\rangle$ states can be used as the spin states $s_n$.
The coupling constant $J_{nm}$ is determined by the dipole-dipole
interaction between molecules in different lattice sites. The
rotational excitation $|g \rangle \rightarrow | e \rangle$ thus
leads to spin waves of the Ising model. In order to simulate more
complicated spin models, such as Eq.~(\ref{H heisenberg}), it is
necessary to use molecules with more complex structure
\cite{book-chapter,N-A-spin-crystal}.

\subsection{Bosons, Fermions and Paulions} \label{s-s-mappings}

The Hamiltonians presented in section \ref{s-s-lattice models}
describe very different physical systems, consisting of particles
with all possible statistics (bosons, fermions, spin matrices and
pseudospins). They are, however, mutually related. Some of these
Hamiltonians can be mapped onto each other, even when the quantum
statistics of mapping and mapped particles are different.

Consider an ensemble of arbitrary two-level systems on a lattice.
The state of a two-level system in lattice site $n$ can be
characterized by the operators $\hat p_n^\dag$ and $\hat p_n$
describing, respectively, the creation and destruction of the
excited state. Operators for different lattice sites necessarily
commute, as they act on different variables. On the other hand,
the same site cannot accommodate more than one excitation, which
is a feature of Fermi statistics. The commutation relations for
the operators $\hat p_n$ are
\begin{eqnarray}\label{paulion commutations}
\begin{array}{c}
\hat p_n \hat p_m^\dag - \hat p_m^\dag \hat p_n = 0 \hskip 0.5 cm
(n \neq m)\\

\\

\hat p_n \hat p_n^\dag + \hat p_n^\dag \hat p_n = 1.
\end{array}
\end{eqnarray}

The operators with such ``mixed'' statistics are called paulions
(in condensed matter physics \cite{kaplan}) or hard-core bosons
(in atomic physics \cite{girardeau}). Using the $\hat
p$-operators, we can write the following general Hamiltonian
\begin{equation} \label{paulionic hamiltonian} \hat
H_{p} = E_0 \sum\limits_{n} \hat p_{n}^\dag \hat p_n +
{\sum\limits_{n,m}}^{'} t_{n,m} \hat p_{n}^\dag \hat p_{m} +
{\sum\limits_{n,m}}^{'} V_{n,m} \hat p_{n}^\dag \hat p_{m}^\dag
\hat p_{n} \hat p_{m}
\end{equation}
which is identical to the hard-core Hamiltonian (\ref{H t-V}) with
the constraint (\ref{H t-V constraint}) absorbed into the
statistical properties of paulions (\ref{paulion commutations}).
The prime over the sum symbols indicates that $n\neq m$. For
$V=0$, i.e. when the density-density correlations are absent, the
Hamiltonian (\ref{paulionic hamiltonian}) is identical to the
Ising Hamiltonian (\ref{H ising}) with $h=0$. This Hamiltonian is
also used to model Frenkel excitons in solid state molecular
crystals \cite{agranovich-book}.

Any unitary transformation preserves the commutation properties of
the bosonic and fermionic operators. For example, the Fourier
transform of bosons must produce bosons. Therefore, many articles
have been devoted to mapping paulions onto ``effective'' particles
with bosonic or fermionic statistics\footnote{We note that similar
transformations exist also for spin matrices $\sigma_n^+ =
\sigma_n^x + i \sigma_n^y$, $\sigma_n^- = \sigma_n^x - i
\sigma_n^y$ and $\sigma_n^z$. They can be mapped onto fermionic
operators by the Jordan--Wigner transformation
\cite{jordan-wigner}, and onto bosonic operators by the
Holstein--Primakoff transformation \cite{holstein-primakoff}.}.
Girardeau showed \cite{girardeau} that in 1D the many-body wave
function of hard-core bosons -- paulions -- corresponds (up to a
sign) to a many-body wave function of a gas of fictitious
non-interacting spin-less fermions. The exact mapping was proposed
later by Chestnut and Suna for a 1D system in the nearest neighbor
approximation \cite{chestnut-suna}. This is a variant of the
Jordan--Wigner transformation \cite{jordan-wigner}, and it works
very well in 1D. However, in higher dimensions it may cause
problems: the effective fermionic operators are non-local, i.e.
the expression for the fermionic operators in site $n$ is
dependent on the occupation numbers at other sites. That is why
this transformation is not as effective in dimensions higher than
one. Another possibility is to use the Agranovich--Toshich
transformation \cite{agranovich-toshich}, which expresses
$\hat{p}$-operators through an infinite series of bosonic
operators $\hat{b}_n$ and $\hat{b}_n^\dag$, in a way ensuring that
for any number of bosons per lattice site $\hat{N}_n^{(b)} =
\hat{b}_n^\dag \hat{b}_n = 0,1,2,3...$ the eigenvalues of the
number operator for paulions $\hat{N}_n = \hat{p}_n^\dag
\hat{p}_n$ are only 1 or 0, so that the unphysical states with
$\hat {N}_n > 1$ do not occur. Then the paulionic Hamiltonian
(\ref{paulionic hamiltonian}) reduces to a sum of terms consisting
of the same number of creation and annihilation bosonic operators,
which we schematically denote by $\hat H_\nu =
\sum\limits_{\{n,m\}} \alpha_\nu \hat b_{n_1}^\dag ... \hat
b_{n_\nu}^\dag \hat b_{m_1} ... \hat b_{m_\nu}$; $\nu =
1...\infty$. Keeping the pairwise interactions only and assuming
that $|t| \ll E_0$, one obtains:
\begin{eqnarray}
\hat{H} = E_0 \sum\limits_{n}\hat{b}_n^\dag \hat{b}_n +
\sum\limits_{n} t_{n,m} \hat{b}_n^\dag \hat{b}_{m} +
 {\sum\limits_{n,m}}^{'} V_{n,m} \hat b_{n}^\dag \hat
b_{m}^\dag \hat b_{n} \hat b_{m} + \hat{H}_{\rm kin},
\end{eqnarray}

\noindent
where
\begin{eqnarray}
\hat{H}_{\rm kin}
= - 2 E_0 \sum\limits_{n=1}^{\cal N} \hat{b}_n^\dag \hat{b}_n^\dag
\hat{b}_n \hat{b}_n.
\end{eqnarray}

\noindent Thus, the Agranovich--Toshich transformation reduces the
pauilonic (hard-core boson) Hamiltonian to a Hamiltonian
describing a gas of bosons interacting via delta-like pairwise
interaction with the strength $2E_0$. Higher-order corrections to
the interaction energy can be obtained by including the omitted
terms. The Hamiltonian $\hat H_{\rm kin}$ describes the kinematic
interaction, which we discuss in section \ref{s-s-s-kinematic}. In
contrast to the Wigner--Jordan transformation, this approach is
very effective in 3D, where $\hat{H}_{\rm kin}$ can be treated as
a perturbation. For 2D systems, the delta-like scattering with the
magnitude $2E_0$, which is the largest energy scale of the
problem, is very strong and does not allow for a perturbative
treatment \cite{ML}. In 1D, the effect of scattering is even
stronger and the transformation to bosons makes no sense at all.
To summarize, in 1D paulions are well described by a gas of
fermions, in 3D by bosons, in 2D they are something in between.
Finally, we note that in principle it is not necessary to use any
of these transformations; one can work directly with paulions
taking into account their commutation relations (\ref{paulion
commutations}).

\section{Rotational Frenkel excitons}
\label{s-s-s-excitons}

While the majority of research with atoms and molecules on optical
lattices has so far focused on quantum transport of particles in a
lattice potential or quantum simulation of lattice models, such as
the ones described above, the ability to control the structure of
molecules trapped on an optical lattice can also be exploited to
study quantum energy transport and collective excitations
reminiscent of excitons in solid state crystals. In this section,
we discuss the formation of rotational excitons. These excitons
have unique properties and can, in turn, be used as a basis for
quantum simulation of new physical phenomena and new lattice
models that cannot be realized with atoms on an optical lattice.
This is discussed in sections \ref{s-s-nonlinear} --
\ref{s-s-s-holstein polaron}.

\begin{figure}[ht]
\centering
\includegraphics[scale=0.7]{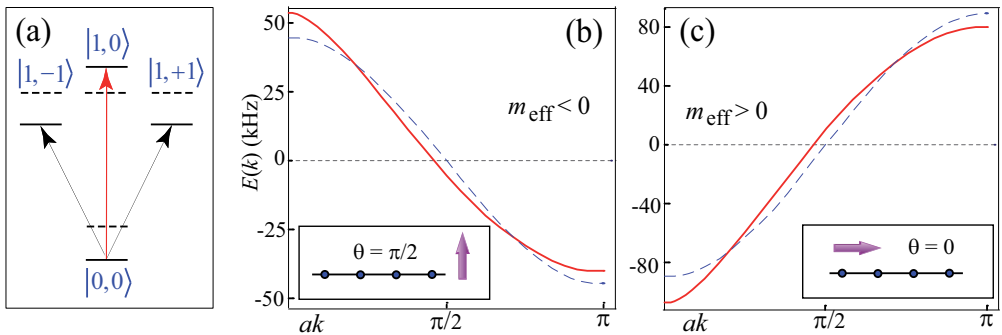}
\caption{Panel (a): Rotational energy levels of a $^1\Sigma$
molecule in the presence of an electric field. The dashed lines
show the positions of the field-free states. Panels (b) and (c):
Energy of rotational excitons in a one-dimensional array of LiCs
molecules trapped on an optical lattice with the lattice constant
$a = 400$ nm in the presence of an electric field of 1~kV/cm. Full
curves -- exact numerical calculation; dashed curves -- analytical
result of Eq.~(\ref{NNA}). The electric field is directed
perpendicular to the array axis (panel b) and parallel to the
array axis (panel c). \label{f-exciton dispersion}}
\end{figure}

Rotational excitation of polar molecules trapped on an optical
lattice in the Mott insulator phase with one molecule per site
gives rise to the formation of rotational Frenkel excitons
\cite{disordered-paper}. The Frenkel exciton is a charge-less
quasiparticle, which describes the excitation transfer in
molecular crystals \cite{Ya.-I.-Frenkel}. Here, we consider the
transition between the absolute ground state $| g \rangle$ of
trapped polar molecules and their first rotational excited state.
We assume that the molecules reside in the ground vibrational
state of the ground electronic state $^1 \Sigma$. In the presence
of a dc electric field ${\cal E}_f$, the three-fold degeneracy of
the rotationally excited state $| N=1, M_N\rangle$ is lifted and
the state with the projection $M_N = 0$ of the total angular
momentum is detuned from the states with $M_N = \pm 1$ (Figure
\ref{f-exciton dispersion}a). We assume that the detuning is large
enough so that the latter can be disregarded and consider the
isolated two-level system of the state $|g \rangle$ and the
rotational excited state with $M_N=0$, denoted by $| e \rangle$.

The molecular states are the eigenstates of the Hamiltonian
\begin{equation}
\hat H_{n}^{\rm (mol)} = B_e \hat N_n^2 - \hat {\bf  d}_n \cdot
\vec{\cal E}_f,
\end{equation}
where $B_e$ is the
rotational constant, $\hat N_n$ is the operator of the angular
momentum, and $\hat {\bf d}_n$ is the dipole
moment of the molecule in site $n$. The field-dressed states are
linear combinations of the field-free rotational states
\begin{equation}\label{|g> and |e>}
\begin{array}{c}
| g \rangle = \sum\limits_N \alpha_N({\cal E}_f)\ | N, M_N = 0
\rangle,\\

\\

| e \rangle = \sum\limits_N \beta_N({\cal E}_f)\ | N, M_N = 0
\rangle.\\
\end{array}
\end{equation}
The coefficients $\alpha_N$ and $\beta_N$ are determined by the
electric field strength ${\cal E}_f$. In the limit ${\cal E}_f
\rightarrow 0$, the states $| g \rangle$ and $| e \rangle$ become,
respectively, $| N=0, M_N =0 \rangle$ and $| N=1, M_N=0 \rangle$.

It is convenient to introduce the {\it transition} operators $\hat
P_n$ defined by the equations $\hat P_n^\dag | g_m \rangle =
\delta_{nm} | e_n \rangle$ and $\hat P_n| e_m \rangle =
\delta_{nm} | g_n \rangle$. As discussed in section
\ref{s-s-mappings}, these operators describe hard-core bosons, or
paulions, and the excitonic Hamiltonian describes a hard-core
boson gas with long-range interactions, which is equivalent to the
Ising model. The rotational excitons can thus be mapped onto other
systems described by the {\it t--V} model, or the Ising model.

Using the $\hat P_n$-operators, the total Hamiltonian for
$\cal{N}$ molecules on an optical lattice can be written as
\cite{agranovich-book}
\begin{eqnarray}
\hat H_{\rm tot} = \hat H_{\rm exc} + \hat H_{\rm dyn} + \hat
H_{\rm n-c},
\end{eqnarray}
where the first term accounts for the excitation transfer between
lattice sites
\begin{eqnarray}\label{Hexc}
\hat H_{\rm exc} = \sum\limits_{n} E_0 \hat P_{n}^\dag \hat P_{n} +
{\sum\limits_{n,m}}^{'} J(n-m) \hat P_{n}^\dag \hat P_{m}
\end{eqnarray}
with the constant $J(n-m)$ determined by the matrix elements of
the inter-molecular excitation transfer due to dipole-dipole
interaction
\begin{eqnarray}
 J(n-m) = \langle e_n g_m | \hat V_{dd} (n-m) | g_n e_m \rangle.
\end{eqnarray}
The second term describes non-linear interactions between
excitons:
\begin{eqnarray}\label{Hdyn}
\displaystyle \hat H_{\rm dyn} = \frac{1}{2}
{\sum\limits_{n,m}}^{'} D(n-m) \hat P_{n }^\dag \hat P_{m}^\dag
\hat P_{n} \hat P_{m},
\end{eqnarray}
with the interaction constant $D(n-m)$ determined by
\begin{eqnarray}\label{D(n)}
\nonumber
 D(n-m) = \langle e_n e_m | \hat V_{dd} (n-m) | e_n
e_m \rangle + \hspace{4.cm}
\\
\langle g_n g_m | \hat V_{dd} (n-m) | g_n g_m
\rangle -
2 \langle e_n g_m | \hat V_{dd} (n-m) | e_n g_m \rangle.
\end{eqnarray}
Finally, $\hat H_{\rm n-c}$ contains terms that do not conserve
the number of molecular excitations in the system:
\begin{eqnarray}
 \hat H_{n-c} = {\sum\limits_{n,m}}^{'} \langle e_n
e_m | \hat V_{dd}(n-m) | e_n g_m \rangle \Bigl( \hat P_{n} + \hat
P_{n}^\dag \Bigr) + \hspace{3.cm}
\nonumber
\\
\frac{1}{2} {\sum\limits_{n,m}}^{'} \langle e_n e_m | \hat
V_{dd}(n-m) | g_n g_m \rangle \Bigl( \hat P_{n} \hat P_{m} + \hat
P_{n}^\dag \hat P_{m}^\dag \Bigr) +\hspace{3.cm} \nonumber
\\
 + {\sum\limits_{n,m}}^{'} \Bigl[\langle g_n g_m |
\hat V_{dd}(n-m) | g_n e_m \rangle - \hspace{4.cm} \nonumber
\\
\langle e_n e_m | \hat V_{dd}(n-m) | e_n g_m \rangle \Bigr] \Bigl(
\hat P_n^\dag \hat P_{n} \hat P_{m} + \hat P_{n}^\dag \hat
P_{m}^\dag \hat P_n \Bigr). \hspace{2.cm} \label{H-n-c}
\end{eqnarray}

If $| g \rangle$ and $| e \rangle$ are states of well-defined
parity (such as, for example, the rotational states at zero
electric field), the matrix elements $D(n - m)$ as well as the
linear and cubic terms in Eq.~(\ref{H-n-c}) must vanish. This is
the case for molecular solids, such as anthracene and naphtalene,
which are often considered as prototype systems for (electronic)
Frenkel excitons \footnote{The terms of the kind $\hat P \hat P$
and $\hat P^\dag \hat P^\dag$, which do not conserve the number of
particles, do not vanish in natural molecular crystals with the
inversion symmetry. The usual model neglecting these terms is
known as the Heitler--London approximation \cite{agranovich-book}.
These terms can be accounted for as corrections, whose magnitude
depends on the ratio between the dipole-dipole interaction matrix
element $J(a)$ and the excitation energy of the molecules $E_0$,
as well as on the structure of the molecules
\cite{beyond-Heit-Lon}.}. For an ensemble of molecules in an
optical lattice, the magnitude of these terms can be tuned by
applying an external electric field, which breaks the inversion
symmetry (parity) of the molecules. To our knowledge, the role of
the odd-$\hat P$ terms in Eq.~(\ref{H-n-c}) has not been
explicitly studied and remains an interesting open problem. The
possibility to tune these terms by varying the field dressing of
the states $|g\rangle$ and $|e\rangle$ may lead to new interesting
phenomena. In what follows, we assume that the effect of $\hat
H_{\rm n-c}$ is small and neglect this term.

Due to the translational invariance of the optical lattice, the
Hamiltonian $\hat H_{\rm exc}$ can be diagonalized by the
transformation:
\begin{equation}
\begin{array}{c}
\hat P^\dag_n = \frac{1}{\sqrt{\cal N}} \sum\limits_q e^{-iqn}
\hat P^\dag(q),\\

\\

\hat P_n = \frac{1}{\sqrt{\cal N}} \sum\limits_q e^{iqn} \hat
P(q),\\
\end{array}
\end{equation}
to yield
\begin{eqnarray}
 \hat H_{\rm exc} = \sum\limits_q E(q)\hat P^\dag(q)\hat P(q),
\end{eqnarray}
where $\hat P^\dag(q)$ and $\hat P(q)$ describe the Bloch plane
waves, $q$ is the (linear) momentum, and $E(q) = E_0 +
\sum\limits_n J(n) e^{iqn}$ is the exciton energy forming a
quasi-continuous band. For ${\cal N} \gg 1$, we can consider the
wave vector $q$ and the energy $E(q)$ to be continuous variables.

In the nearest neighbor approximation,
\begin{eqnarray}
E_{\rm NNA}(q) = E_0 + 2J \cos aq
\label{NNA}
\end{eqnarray}
and
\begin{eqnarray}
 J = \langle e_n g_{n+1}|
\hat V_{dd}(a)| g_n e_{n+1} \rangle,
\end{eqnarray}
where $a$ is the lattice constant. The energies are shown in
Figure~\ref{f-exciton dispersion} for different angles $\theta$
between the one-dimensional molecular ensemble and the electric
field ${\cal E}_f$ (which determines the sign of $J$ and,
consequently, the shape of the dispersion curve). In the nearest
neighbor approximation, $\Delta = 4J$ is the exciton bandwidth.
The magnitude of $J$ is a quantitative measure of the collective
excitation effects (in particular, $h/J$ is the timescale of the
excitation transfer between molecules in adjacent lattice sites).
Note that $J=0$ at $\theta = \theta^* \equiv \arccos
(1/\sqrt{3})$: at this angle the dipole-dipole interaction
vanishes, and (in the dipole approximation) the molecules become
decoupled. When this happens, no excitation transfer dynamics can
occur. For polar molecules with permanent dipole moments of a few
Debye, the values of $J$ can be tens of kHz (for LiCs on an
optical lattice with $a=400$~nm, $J \sim 25$~kHz).

For small $q$, where the dispersion is parabolic, we can introduce
the effective mass $m_{\rm eff} = -\hbar^2 / J a^2$. For $\theta >
\theta^*$ ($\theta < \theta^*$) the constant $J$ is positive
(negative), and, consequently, the effective mass is negative
(positive). Quasiparticles with negative effective mass have
counterintuitive propagation properties: they move in the
direction opposite to their wave vector (negative refraction
\cite{negative-refraction}). Thus, by changing the angle $\theta$
it may be possible to tune the excitons from the regime of normal
propagation to the regime of negative refraction.

The possibility of tuning the parameters of the exciton
Hamiltonians (\ref{Hexc}), (\ref{Hdyn}) and (\ref{H-n-c}) in an
ensemble of polar molecules on an optical lattice opens the
possibility to access new regimes of Frenkel exciton physics that
cannot be observed in solid state crystals. Here, we discuss a few
examples that we presented in our recent papers
\cite{disordered-paper}, \cite{polaron-paper},
\cite{biexciton-paper}.

\section{Nonlinear interactions of Frenkel excitons}
\label{s-s-nonlinear}

In this section we study non-linear interactions between
rotational Frenkel excitons and show that they can be dynamically
tuned by an external electric field. This is important for
applications of excitons in quantum information processing
\cite{rabl-zoller-paper}, where excitons can be used as qubits. In
particular, we consider the interplay of the dynamical
interactions arising from Eq.~(\ref{Hdyn}) and the kinematic
interaction arising from the hard-core boson nature of the exciton
operators and show that the relative importance of these two
interactions can be tuned.

\subsection{Dynamical and kinematic interactions}
\label{s-s-s-kinematic}

There are two types of non-linear interactions for Frenkel
excitons. The first is the {\it dynamical} interaction given by
Eq.~(\ref{Hdyn}). This interaction is determined by the matrix
elements $D(n)$ of the dipole-dipole interaction (see
Eq.~(\ref{D(n)})). In the wave vector representation, the
dynamical interaction describes wave vector conserving scattering
between two excitons with the exchange of momentum $q$
\begin{equation}\label{dynamical wave vector}
\hat{H}_{\rm dyn} = \frac{1}{\cal N} \sum\limits_{k_1, k_2, q} D(q)
\hat{P}^\dag(k_1+q) \hat{P}^\dag(k_2-q) \hat{P}(k_1) \hat{P}(k_2).
\end{equation}
where $D(q) = \sum\limits_{n=1}^{\cal N} D(n) e^{-i qn}$.

The second interaction mechanism originates from the hard-core
repulsion of molecular excitations. Since excitons are paulions,
or hard-core bosons with long-range interactions, they effectively
repel each other when placed in the same lattice site. As a
result, the direct products of $\hat{P}(k)$-operators are not the
eigenstates of the exciton Hamiltonian $\hat{H}_{\rm exc}$
(\ref{Hexc}), as they would be for bosons or fermions. The
commutation relations for paulions (\ref{paulion commutations})
yield for a two-exciton state $| \Phi(k_1,k_2) \rangle = |
\hat{P}^\dag(k_1) \hat{P}^\dag(k_2) \rangle$ the following
equation:
\begin{equation}\label{kinematic}
\begin{array}{c}
\hat{H}_{\rm exc}\ | \Phi(k_1,k_2) \rangle = [E(k_1) + E(k_2)]\  |
\Phi(k_1,k_2) \rangle +\\

\\

+ \displaystyle \frac{1}{\cal N} \sum\limits_{{q_1,
q_2}\atop{q_1+q_2 = k_1 + k_2}} [E(q_1) + E(q_2)]\ | \Phi(q_1,q_2)
\rangle.\\
\end{array}
\end{equation}
The second term on the right-hand side may be interpreted as total
wave vector conserving scattering between plane-wave-like
one-exciton states $| \hat{P}^\dag(q_i) \rangle$ \cite{kaplan}.
This type of scattering results from the kinematic interaction
\cite{dyson}. This interaction can also be included as a
perturbative term $\hat H_{\rm kin}$ in the total Hamiltonian for
the effective bosons introduced via the Agranovich--Toshich
transformation (section \ref{s-s-mappings}). The non-linear
properties of excitons (in the two-body interaction approximation)
are determined by the balance between the dynamical and kinematic
interactions.

\subsection{External field control of exciton--exciton
interactions} \label{s-s-s-exc-exc interaction}

In the presence of an external electric field and with account of
the dynamical interaction, the rotational Frenkel excitons are
described by the Hamiltonian $\hat H_{\rm nl} = \hat H_{\rm exc} +
\hat H_{\rm dyn}$ given in Eqs.~(\ref{Hexc}) and (\ref{Hdyn}). In
the nearest neighbor approximation $\hat H_{\rm exc}$ and $\hat
H_{\rm dyn}$ are parametrized by two constants: $D = D(a)$ and $J
= J(a)$. Since $J$ determines the propagation properties of
excitons, it is implicitly related to the strength of the
exciton--exciton kinematic interaction: roughly speaking, the
strength of repulsive interaction (scattering) between two
excitons at the same lattice site is proportional to $|J|$. The
constant $D$ describes the dynamical interaction. The paulionic
corrections do not affect $\hat H_{\rm dyn}$ so one can use the
bosonic commutation relations for the $\hat P$-operators in this
term, both in the site and wave vector representations.

\begin{figure}[ht]
\centering
\includegraphics[scale=0.7]{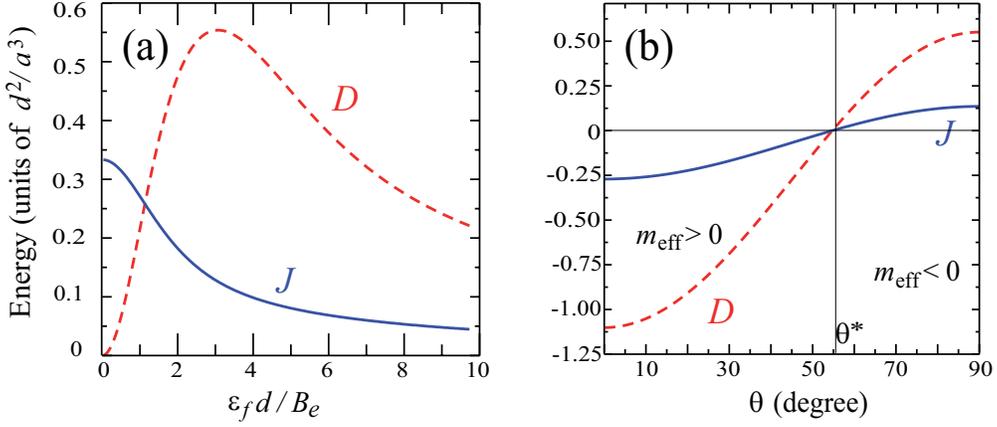}
\caption{(a) $D$ and $J$ as functions of electric field magnitude
at a fixed angle $\theta = 90^o$. (b) $D$ and $J$ as functions of
$\theta$ at a fixed electric field ${\cal E}_f = 6$~kV/cm. The
calculations are for a 1D ensemble of LiCs molecules separated by
$a$=400~nm; $d^2/a^3$ = 72 kHz. \label{f-D and J}}
\end{figure}

Both $D$ and $J$ can be controlled by varying the magnitude and
orientation of the electric field (see Figure \ref{f-D and J}).
Panel (a) of Figure \ref{f-D and J} shows the dependence of $D$
and $J$ for a one-dimensional array of LiCs molecules on the field
magnitude at a fixed angle between the array axis and the
direction of the electric field. Note that $D$ vanishes at zero
electric field (see Eq.~(\ref{D(n)})). For ${\cal E}_f = 0$, the
field-dressed states $|g \rangle$ and $|e\rangle$ reduce
adiabatically to the pure rotational states: $| g \rangle \to |
N=0, M_N =0 \rangle$ and $| e \rangle \to | N=1, M_N=0 \rangle$,
and the matrix elements determining $D$ vanish according to the
selection rules. Panel (b) shows the dependence of $D$ and $J$ on
the angle $\theta$ for a fixed field magnitude. Note that both $D$
and $J$ vanish at $\theta^\ast = \arccos (1/\sqrt{3})$.

Figure \ref{f-D and J}b shows that the matrix elements $D$ and $J$
have the same sign, independent of $\theta$. As discussed in
section~\ref{s-s-s-excitons}, the sign of the effective mass is
always opposite to the sign of $J$, and, consequently, the sign of
$D$. The constant $D$ determines the dynamical interaction
potential (\ref{Hdyn}). Due to the linearity of the
Schr\"{o}dinger equation, a positive potential is attractive for a
particle with negative mass and a negative potential is attractive
for a particle with positive mass \cite{disordered-paper}.
Therefore, the dynamical interaction in this system is always
attractive \cite{biexciton-paper}. The kinematic interaction is
always repulsive because it arises as a consequence of the
hard-core boson nature of the excitation operators.

Figure~\ref{f-D and J} shows that the values of $D$ and $J$ can be
tuned in a wide range of magnitudes. For the chosen example of
LiCs molecules on an optical lattice with $a = 400$ nm, $D$ can be
varied in the interval from -80 kHz to 40 kHz and $J$ from  -50
kHz to 25 kHz. Their relative magnitudes determine the relative
contributions of the kinematic and dynamical interactions. By
changing the ratio between $D$ and $J$, one can explore the
interaction regimes dominated by the different interactions. In
particular, prevailing attraction may result in the formation of
bound two-exciton complexes known as biexcitons
\cite{biexciton-paper}. They form in the low dimensional systems
when $|D| > 2|J|$ \cite{vektaris}. In 1D they appear as a single
state band split form the continuum of the two-particle states. As
a consequence of the correlation between the signs of $D$ and the
effective mass, the biexciton state appears below (above) the
two-particle continuum for negative (positive) $D$, and can be
continuously tuned between these two positions by varying the
angle between the electric field and the intermolecular axis. The
properties of biexcitons can be engineered by varying the ratio
$D/2J$ by tuning the magnitude of the electric field. In
particular, for an ensemble of LiCs molecules, the biexciton
begins to appear at ${\cal E}_f \approx 3.6$~kV/cm, when $D=2J$.
With the increase of the electric field its binding energy (and,
correspondingly, its splitting from the continuum states)
increases, and the wave function shrinks, so that at $D \gg 2J$ a
biexciton is a strongly correlated state of two molecular
excitations separated by one or two lattice constants. The
formation of a biexciton thus resembles the association of a
molecule by combining two atoms.

When $D=0$, the dynamical interaction vanishes, and the kinematic
interaction dominates. It should be mentioned that the effect of
the kinematic interactions on Frenkel exciton dynamics has not yet
been observed in experiments. To this end, it would be useful to
find a mechanism for tuning the kinematic interaction as well.  We
explore this in the following section.

\subsection{Suppression of kinematic interaction}

Although the kinematic interaction is inherent to molecular
crystals as an intrinsic consequence of the exciton operator
statistics, it may be possible to generate excitations that do not
experience kinematic interactions. To find the conditions for such
excitations, we neglect the dynamical interactions and look for
solutions of the two-particle Schr{\" o}dinger equation
\begin{eqnarray}
\hat H_{\rm exc} |\Psi_K \rangle = E_K | \Psi_K \rangle
\end{eqnarray}
in the form
\begin{eqnarray}
| \Psi_K \rangle = \sum\limits_k C_K(k) \hat P^\dag(K/2 + k) \hat
P^\dag(K/2 - k) | \rm{vac} \rangle, \label{two-particle}
\end{eqnarray}
where $k = (k_1 - k_2)/2$ is the relative wave vector of two
interacting excitons, and $K = k_1 + k_2$. The expansion
coefficients in Eq.~(\ref{two-particle}) satisfy the following
equation
\begin{equation}\label{2body disp eq}
\biggl( \varepsilon_K(k) - E_K \biggr) C_K(k) =
\frac{\varepsilon_K(k)}{\cal N} \sum\limits_{k^{'}} C_K(k^{'}),
\end{equation}
where $\varepsilon_K(k) = E(K/2 - k) + E(K/2 + k)$ is the total
energy of two interacting excitons. The two-particle amplitude in
the site representation, which depends on the relative distance $r
= n-m $ between two molecular excitations, is $C_K(r) = 1/{\cal N}
\sum\limits_k e^{ikr} C_K(k)$. Since $C_K(r=0) = 0$, the
right-hand side of Eq.~(\ref{2body disp eq}) vanishes. Therefore,
Eq.~(\ref{2body disp eq}) can be satisfied only if
\begin{equation}\label{kin suppress}
\varepsilon_K(k) = {\rm const} = E_K.
\end{equation}

The corresponding wave function is $({\cal N} - 1)$-time
degenerate ($n$-degeneracy):
\begin{equation}
| \Psi_{K}^{(n)} \rangle = \frac{1}{\sqrt{\cal N}} \sum\limits_k
B_{K}(n) e^{ikn} \hat P^\dag (K/2 + k) \hat P^\dag (K/2 - k) |
\rm{vac} \rangle,
\end{equation}
where the quantum number $n$ determines the fixed distance $r =
na$ between the excited molecules: $n = 1, 2, ... , {\cal N}-1$.
These states describe two correlated excitations that do not
experience the kinematic interaction.

The key requirement (\ref{kin suppress}) -- the $k$-independence
of the two-particle energy $\varepsilon_K(k)$ -- is however not
easy to satisfy. One possibility is to consider a specific choice
of $K$. In the nearest neighbor approximation, when $E(q) = E_0 +
2J \cos aq$, the two-particle energy written in terms of $K$ and
$k$ is $\varepsilon_K(k) = 2E_0 + 4J \cos (ak) \cos (aK/2)$. It
reduces to a constant if $K = \pi/a$. Thus, with the accuracy up
to corrections coming from the interactions beyond the nearest
neighbors and quadrupole interactions, such a pair immune to the
kinematic interaction can be produced starting from a pair of
excitons with $q_1 + q_2 = \pi/a$. Combined with the possibility
to control the dynamical interaction strength by the electric
field magnitude, this provides a system in which both attractive
and repulsive interactions between quasiparticles can be
independently tuned.

Another possibility to produce a two-particle state with the
energy independent of $k$ can be realized using the three-level
structure of the first rotationally excited state of $^1\Sigma$
molecules (see Figure \ref{f-exciton dispersion}a). At low
electric fields, when the energy separation between the levels $|
N=1, M=0 \rangle$ and $| N=1, M=\pm 1 \rangle$ is small, the
different excitation branches may be mixed, which corresponds to
configuration mixing in solid state molecular crystals
\cite{agranovich-book}. Due to the symmetry properties of the
molecular system considered here, the state with $M=0$ is
decoupled from the states with $M = \pm 1$ if the electric field
is parallel or perpendicular to the molecular ensemble. For other
angles $\theta$, the collective excitations give rise to three
exciton modes which we denote $\alpha$, $\beta$ and $\gamma$,
containing contributions from all three molecular transitions
\cite{disordered-paper}:
\begin{equation}\label{PM(k)}
\hat P_{\rho = \alpha,\beta,\gamma}^\dag(k) = \sum_{M=-1,0,1}
u_M^\rho(k) \hat P_M^\dag(k),
\end{equation}
where $\hat P_M^\dag(k) = (1/\sqrt{\cal N}) \sum_n e^{i kn} \hat
P_{n,M}^\dag$ are the exciton operators corresponding to the
molecular transition to the excited state with projection $M =
M_N$. Operators (\ref{PM(k)}) diagonalize the four-level excitonic
Hamiltonian yielding
\begin{equation}
\hat H_{\rm exc}^{\{M\}} = \sum\limits_{k; \rho =
\alpha,\beta,\gamma} E_\rho(k) P_{\rho}^\dag(k) P_{\rho}(k).
\end{equation}

At high electric field, the $\gamma$-mode is split from the other
two modes, leading to the isolated one-band exciton discussed
throughout this article. However, the other two branches
correspond to the excitations of the degenerate molecular states
and remain mixed. Their $k=0$ state can be accessed by microwave
field with circular polarization. The non-zero $k$ states can be
probed by Raman transitions combining photons with linear and
circular polarization. In the limit of high electric fields, the
energies of these excitons can be written as
\begin{equation}\label{Ealpha, Ebeta}
\begin{array}{c}
E_\alpha(k; \theta) = E_1({\cal E}_f) + 2 |\alpha_{N=0}({\cal
E}_f)|^2 j(k)\biggl( \cos^2
\theta - 2/3\biggr),\\

\\

E_\beta(k) = E_1({\cal E}_f) + 2 |\alpha_{N=0}({\cal E}_f)|^2
j(k)/3,
\end{array}
\end{equation}
where $E_1({\cal E}_f)$ is the transition energy between the
field-dressed ground state and the field-dressed excited state
with $M = \pm 1$; $j(k) = (d^2/a^3) \sum_{n=1}^{{\cal N}/2} \cos
(akm)/m^3$, and $\alpha_N({\cal E}_f)$ are defined in
Eq.~(\ref{|g> and |e>}). The kinematic interaction in the presence
of several branches is more complicated, but it can be shown that
the condition (\ref{kin suppress}) also applies to the
multi-branch problems. It can be seen that, at a particular angle
$\theta = \arccos \sqrt{1/3}$,
\begin{equation}\label{33}
E_\alpha(k; \theta) + E_\beta(k)  = 2E_1({\cal E}_f) = {\rm
const},
\end{equation}
so the kinematic interaction must be absent. Eq.~(\ref{33}) is
satisfied for arbitrary values of $K$. We note that for the
$\gamma$-branch, the matrix elements of the dipole-dipole
interaction vanish at $\theta = \theta^\ast$ so the $\gamma$-state
becomes dispersion-less. However, the matrix elements giving rise
to the other two exciton states remain non-zero at this angle, see
Eqs.~(\ref{Ealpha, Ebeta}).  It is the cancellation of the
dispersions of $E_{\alpha}(k)$ and $E_{\beta}(k)$ that leads to
the suppression of the kinematic interaction.

\section{Rotational excitons in a tunable disordered potential}

Dynamics of quantum particles in disordered potential has been
extensively studied in relation to Anderson localization
\cite{Anderson-loc-review}, propagation of light
\cite{disorder-light} and particles \cite{disorder-matter-waves}
through disordered media, phase transitions between insulating and
conducting states \cite{isulator-conductor-in-solid}, to name a
few examples. As a result of these studies, many properties of
disordered systems, such as the role of the density of states or
the correlation functions are well understood. At the same time,
there are still many problems that remain open. A few examples
include the role of disorder in the transition to the glassy state
\cite{glass}, transition from a superconductor to insulator with
increasing disorder \cite{supercond-insulator}, and the
counterintuitive behavior of conductivity in some disordered
quasi-crystals \cite{disorder-in-quasicrystals}.

In this section we show that ultracold molecules trapped on an
optical lattice offer the possibility to study quasiparticles in
the presence of dynamically tunable disordered potential. We do
not consider the effects of unavoidable natural disorder, such as
unoccupied lattice sites, lattice potential inhomogeneity or
fluctuations of the electric field. Always present in experiments
with atoms and molecules on optical lattices, the natural disorder
induces localization and decoherence of excitons leading to
homogeneous broadening of exciton dispersion curves. We assume
that the effects of natural disorder can be reduced to a small
fraction of the exciton bandwidth. Here, we consider the
possibility of applying an external disordered potential that
could be varied to allow the observation of real-time dynamics of
disorder-induced phenomena.

\subsection{Optical lattice with tunable disorder}

There are several different models of disorder \cite{zajman}, and
some of them can potentially be realized in an optical lattice
with ultracold molecules. First, suppose that a small fraction of
molecules trapped on the lattice (host molecules) is replaced with
molecules of different kind (impurities). This results in
substitutional disorder. We assume that all impurity molecules are
identical. The impurities break the translational symmetry of the
system and therefore scatter excitons. This scattering is elastic
and modifies the direction of exciton wave propagation, but not
the absolute value of the exciton wave vectors.

An impurity introduced into the molecular crystal in general
differs from the host molecules by the molecular transition
energy, $E_i$, and by the dipole moment, $d_i$, which modifies the
dipole-dipole coupling strength. In the presence of a single
impurity at the lattice site $n=0$, we can write the total
Hamiltonian of the system as
\begin{equation}\label{H exc-imp}
\hat H_{\rm tot} = \hat H_{\rm exc} + V_0 \hat P^\dag_0 \hat P_0 +
\sum\limits_{n \neq 0} \Delta J(n) \biggl( \hat P^\dag_n \hat P_0
+ \hat P^\dag_0 \hat P_n \biggr)
\end{equation}
where $V_0 = E_i - E_0$ and $\Delta J(r)$ is the difference
between the host-host and host-impurity excitation transfer
constants. The exciton-impurity interaction can thus be described
as a sum of a delta-function potential with strength $V_0$ and a
perturbation due to the difference in the dipole moments of the
host and impurity molecules. If the matrix of the operator (\ref{H
exc-imp}) is evaluated in the basis of exciton states in the site
representation, one finds that $V_0$ perturbs the diagonal matrix
elements and $\Delta J$ the off-diagonal matrix elements. Thus,
the constants $V_0$ and $\Delta J(r)$ give rise to diagonal and
off-diagonal disorder.

\begin{figure}[ht]
\centering
\includegraphics[scale=0.7]{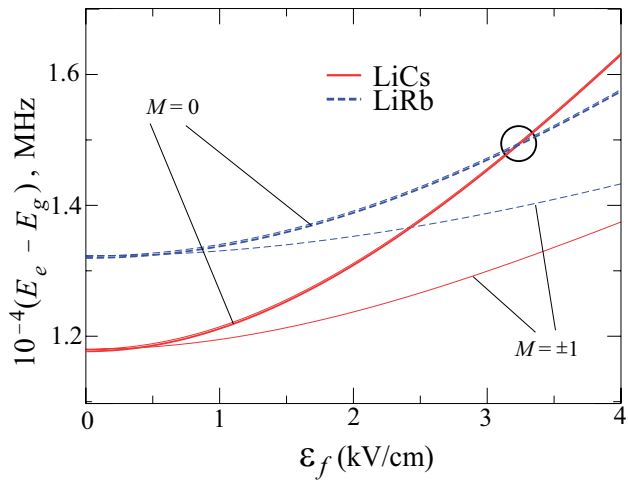}
\caption{Excitation energies of non-interacting molecules in an
electric field for transitions $\ket{N = 0, M=
0}\rightarrow\ket{N= 1,M}$ with $M = 0$ (upper curve) and $M=\pm
1$ (lower curve) vs electric field for LiCs and LiRb.
\label{disorder}}
\end{figure}

For a properly chosen mixture of diatomic molecules, the magnitude
and the sign of $V_0$ can be tuned by an external electric field.
For example, Figure \ref{disorder} shows that this can be achieved
in an array of LiCs molecules doped with LiRb molecules.
Generally, it should be possible to tune $V_0$ from a negative
value to a positive value in a mixture of $^1\Sigma$ diatomic
molecules $AB$ and $CD$, when the dipole moment of $AB$ is greater
and the rotational constant of $AB$ is smaller.

It may also be possible to realize a system with tunable diagonal
disorder in a single-species ensemble of molecules by applying
intense laser beams focused on a small part of the lattice.
Molecules at the focus of the beam must experience larger ac Stark
shifts than the molecules outside the beam focus. These molecules
are equivalent to impurities because their energy level structure
is different. The degree of detuning $V_0$ can be adjusted by
varying the laser field strength. For small fields, it may be
possible to neglect the off-diagonal disorder terms, which
suggests a unique possibility to differentiate between the effects
of diagonal and off-diagonal disorder.

\subsection{Resonant enhancement of exciton--impurity scattering}

The properties of waves with arbitrary dispersion in the presence
of local defects have been studied in Ref.~\cite{IMLifshitz}.
Similar considerations were later applied to excitons in a crystal
with substitutional disorder in Ref.~\cite{dubovsky}, where the
exciton--impurity scattering cross section, which characterizes
their interaction, was derived. The exciton--impurity interactions
lead to the appearance of bound exciton states, which can capture
excitons in solid state crystals \cite{dubovsky-2}. The constants
$V_0$ and $\Delta J$ in Eq.~(\ref{H exc-imp}) determine the
character of the exciton--impurity interactions in natural solids.

We first consider an ensemble of molecules driven by several
focused laser beams with the same strength, so that $\Delta J =
0$, and all impurities are characterized by the same value of
$V_0$. This system may allow for the possibility to explore the
dependence of exciton--impurity scattering cross sections not only
on the exciton wave vector $k$, but also on the impurity
scattering strength $V_0$, which is not possible in conventional
solids. In particular, as we show below, the scattering cross
section can be resonantly enhanced, if the potential produces a
shallow bound state at small values of $V_0$
\cite{disordered-paper}.

For a particle with the parabolic dispersion $E = \hbar^2 q^2 / 2
m_*$ in a $d$-dimensional delta-like potential $\hat H_\delta =
a^d V_0 \delta({\bf r})$ the bound state is split from the
continuum states by the energy $E_b$, which depends on the
dimensionality of the system. The binding energy can be obtained
from the following equations:
\begin{equation}
\begin{array}{c}
\displaystyle \sqrt{\frac{E_b^{\rm (3D)}}{{E}_{\rm loc}}} \arctan
\sqrt{\frac{{E}_{\rm loc}}{E_b^{(3D)}}} = 1 + {\rm sgn} (m_*)
\frac{2{E}_{\rm loc}}{\pi V_0} {\rm \hspace{1.cm} for
\hspace{0.3cm} 3D}\\

\\

\displaystyle \frac{E_b^{\rm (2D)}}{{E}_{\rm loc}} = \left[ \exp\left(
-{\rm sgn} (m_*) \frac{4{E}_{\rm loc}}{\pi V_0} \right) - 1
\right]^{-1} {\rm \hspace{1.cm} for \hspace{0.3cm} 2D} \\

\\

\displaystyle \sqrt{\frac{E_b^{\rm (1D)}}{{E}_{\rm loc}}} =
-\frac{{\rm sgn} (m_*) V_0}{{E}_{\rm loc}} \arctan
\sqrt{\frac{{E}_{\rm loc}}{E_b^{(1D)}}} {\rm \hspace{1.cm} for
\hspace{0.3cm} 1D}\\
\end{array}
\end{equation}
where ${E}_{\rm loc} = \hbar^2 \pi^2 / 2 |m_*| a^2$ is the
localization energy of a particle with the mass $|m_*|$ in a
region of dimension $a$. Note that these equations have
solutions only for negative (positive) $V_0$ and positive
(negative) mass $m_*$.

Figure \ref{f-scattering cross section}a shows the behavior of the
local states as functions of the dimensionless potential strength
$|V_0|/E_{\rm loc}$. In 1D and 2D, an attractive delta-like
potential always produces a bound state, and in 3D only starting
from a finite value: $|V_0| > 2 E_{\rm loc} / \pi$. Resonant
scattering may play an important role if the resonant enhancement
of the scattering cross section at $E_b \to 0$ is reached at a
finite value of $V_0$. For instance, in 1D $E_b^{(1D)} \approx
\pi^2 V_0^2 / 4 E_{\rm loc}$ at vanishing $V_0$, and the
shallowing of the bound state is accompanied by the vanishing of
the scattering potential itself. In turn, in 3D one can expect a
resonant enhancement of the scattering cross section for the
potential $|V_0| \sim 2 E_{\rm loc} / \pi = \pi \hbar^2 / |m_*|
a^2$. Near this potential strength, $E_{b}^{(3D)} \approx E_{\rm
loc} (2/\pi + V_0 {\rm sgn}(m_*) / E_{\rm loc})^2$. In 2D, any
potential $|V_0| < 0.2 E_{\rm loc}$ produces a shallow bound
state, whose energy tends exponentially to zero with vanishing
impurity potential strength: $E_{b}^{\rm (2D)} \approx E_{\rm loc}
\exp[4E_{\rm loc}/\pi V_0 {\rm sgn}(m_*)]$ (we recall that $V_0$
and $m_*$ should have different signs for a bound state to
appear). This exponential dependence leads to efficient resonant
scattering for all $0 < V_0 < 0.2 E_{\rm loc}$: as the potential
strength in 2D exponentially exceeds the kinetic energy of the
scattered wave, then, in contrast to the 1D case, the resonant
scattering cannot be considered as weak, even at vanishingly small
$V_0$. At the same time, this means that for $|V_0| \ll 0.2 E_{\rm
loc}$ the local state is so close to zero, that the potential
scatters resonantly only excitons with $k = 0$, which in fact do
not propagate.

\begin{figure}[ht]
\centering
\includegraphics[scale=0.7]{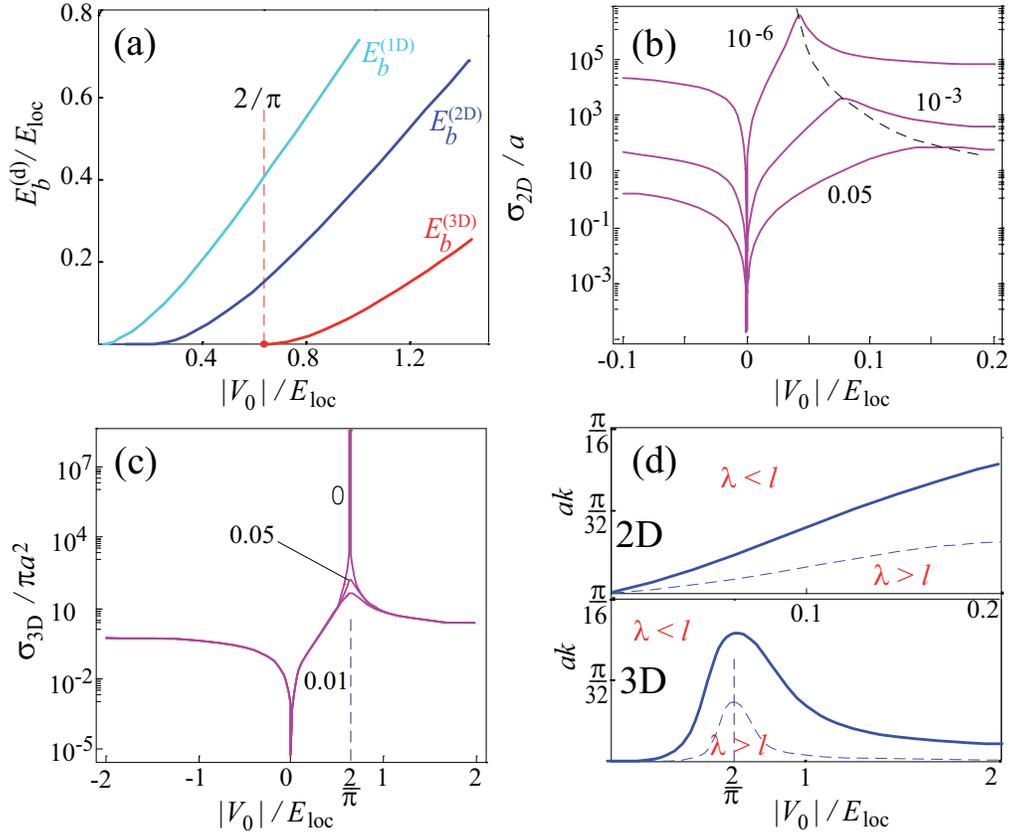}
\caption{(a) Dependence of the bound state energy $E_{b}^{(d)}$ on
potential strength $V_0$ in different dimensions. (b,c) The
exciton--impurity scattering cross sections for 2D (b) and 3D (c)
as functions of the potential strength for different values of
$ak$ (shown near each line). (d) Different propagation regimes in
2D (top) and 3D (bottom) for 1\% of impurities (thick solid line)
and 0.1\% of impurities (thin dashed line). \label{f-scattering
cross section}}
\end{figure}

For small wave vector excitons, $E(k) = (E_0 + 2J) - J a^2 k^2$,
so that $\hbar^2 / m_* = - 2 J a^2$, and $E_{\rm loc} = \pi^2 J$.
Borrowing the results from Ref.\cite{la-rocca}, we write the
scattering cross sections in 2D and 3D for $ak \ll 1$ as explicit
functions of the potential strength $V_0$, and express them as
functions of the bound state energy $E_b$ and the kinetic energy
$T(k) = \hbar^2 k^2 / 2 |m_*| = |J| (ak)^2$ (these expressions
correspond to the limit when both $T(k)$ and $E_b$ are much
smaller than $E_{\rm loc}$):
\begin{equation}\label{sigma 2D 3D}
\begin{array}{c}
\displaystyle \frac{\sigma_{\rm 2D}(k,V_0)}{a} = \frac{\pi^2 /
ak}{\displaystyle \frac{\pi^2}{4} +  \left[ \ln\frac{ak}{\pi} -
\frac{2 E_{\rm loc}}{\pi V_0} {\rm sgn}(m_*)\right]^2} = \frac{4
\pi^2 / ak}{\displaystyle \pi^2 + \left[ \ln \bigl(E_b^{(2D)}/T(k)
\bigr)\right]^2},\\

\\

\displaystyle \frac{\sigma_{\rm 3D}(k,V_0)}{\pi a^2} =
\frac{1}{\displaystyle \left(\frac{ak}{2} \right)^2 + \left[
\left( \frac{ak}{\pi}\right)^2 - \frac{2 E_{\rm loc}}{\pi V_0}{\rm
sgn}(m) -1 \right]^2} = \frac{4 E_{\rm loc}/\pi^2}{E_b^{\rm (3D)}
+ T(k)}.
\end{array}
\end{equation}

Figures \ref{f-scattering cross section}b and \ref{f-scattering
cross section}c show the scattering cross sections for 2D and 3D
(note the divergence of $\sigma_{\rm 3D}$ at $k\to 0, V_0 \to 2
E_{\rm loc}/\pi$). These results demonstrate that by tuning the
potential strength from zero to the critical value ($V_{\rm
cr}^{\rm (2D)} \leq 0.1 E_{\rm loc}$ and $V_{\rm cr}^{\rm (3D)} =
2 E_{\rm loc}/\pi$), one can vary the value of the scattering
cross section for small wave vector excitons by many orders of
magnitude.

The single-impurity scattering cross sections determine the
regimes of exciton propagation in the lattice with a small
admixture of impurities. According to the Ioffe--Regel criterion
\cite{Ioffe-Regel}, excitons with the wavelength $\lambda = 2\pi /
k$ are strongly localized when $\lambda / 2\pi \geq l$ ($l$ is the
elastic mean free path). When $\lambda / 2\pi \ll l$ excitons have
plane-wave-like character, though, in some cases, weak
localization of excitons can be achieved \cite{weakloc}. Excitons
propagate without scattering when $l$ is on the order of the
lattice size.

For the ensemble of polar molecules trapped on an optical lattice,
the elastic mean free path of excitons $l \sim 1 / \sigma c$ for a
given concentration $c$ of impurities can be dynamically tuned by
varying the strength of the electric field that modifies the
disordered potential and, consequently, $\sigma$. This may allow a
possibility to transfer excitons dynamically from the regime of
ballistic propagation to the regimes of weak or strong
localization. Figure \ref{f-scattering cross section}d shows the
$(V_0,k)$--diagram of different propagation regimes for 2D (top)
and 3D (bottom) lattices with the concentration of impurities 1\%.
The solid lines correspond to the condition $l = \lambda / 2\pi$.
Below these lines, the exciton wave length $\lambda$ exceeds $
2\pi l$, which corresponds to the Anderson localization regime.
Above the lines, the states are in general delocalized. The dashed
line is plotted for the impurity concentration 0.1\%. The diagrams
are not complete without another important parameter, the
phase-breaking length $l_\phi$. This length accounts for inelastic
scattering processes. The interference effects leading to weak and
strong localization are only possible if $l \ll l_\phi$
\cite{weakloc}. The phase-breaking length should be calculated for
a given realization of the experimental system with the account of
the major loss channels for rotational excitons.

\subsection{Disorder correlations and localization--delocalization
crossover}
\label{s-s-s-lod deloc}

Consider now a lattice with multiple impurities formed by
molecules of a different type. This leads to significant values of
$\Delta J \neq 0$. Quantum particles in the presence of a random
distribution of scattering centers undergo coherent localization,
and in 1D all states are exponentially localized even in presence
of disorder of arbitrarily small magnitude \cite{1D-loc}. However,
when the disorder potential exhibits short-range correlations,
particular states may become delocalized \cite{deloc-discr}.
Delocalized states may even form a continuous band, if the
correlations are long-range, so that a mobility edge exists
between localized and delocalized states \cite{deloc-contin}.
Studies of correlation-induced delocalization of quantum states
are important for understanding quantum transport in disordered
systems.

The substitutional disorder introduces both the diagonal ($V_0$)
and off-diagonal ($\Delta J$) disorder. As a consequence, $V_0$
and $\Delta J$ are correlated, and the spectrum of a 1D disordered
lattice with a two-molecule mixture must always contain one
delocalized state \cite{tesseri}. The energy of this state is
determined by the relation between $V_0$ and $\Delta J$. The
delocalization occurs because the diagonal and off-diagonal
perturbations compensate one another.

\begin{figure}[ht]
\centering
\includegraphics[scale=0.7]{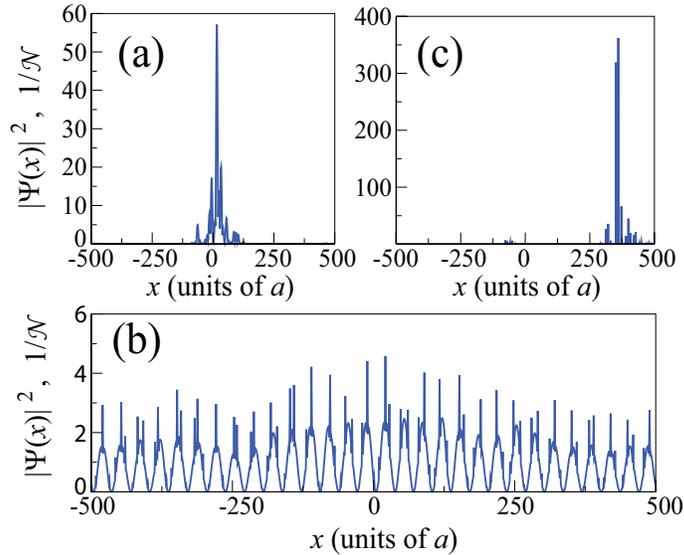}
\caption{Probability density $|\Psi(x)|^2$ describing an exciton
near the top of the energy spectrum for a 1D array of 1000 LiCs
molecules with 10\% homogeneously and randomly distributed LiRb
impurities. Panels correspond to different values of $V_0$: (a)
$V_0 = 0$, (b) $V_0/h = 22$ kHz, and (c) $V_0/h = 100$ kHz. The
difference of the dipole moments of LiCs and LiRb molecules leads
to the value $\Delta J$ = -6.89 kHz. Figure is taken from
Ref.\cite{disordered-paper}. \label{f-loc deloc}}
\end{figure}

As demonstrated by Figure \ref{disorder}, $V_0$ can be tuned by
shifting the rotational levels of host and impurity molecules
simultaneously using a static electric field. In particular, at
the field corresponding to the encircled region, $V_0$ vanishes,
while $\Delta J$ remains finite. Tuning the electric field around
this value allows for the possibility to exploit both the positive
and negative values of $V_0$. Choosing the appropriate value of
$V_0$ can be used to induce the delocalization of any eigenstate
of the system. For states near the origin of the Brillouin zone,
delocalization occurs at $V_0 \approx -4 \Delta J$
\cite{disordered-paper}. Figure \ref{f-loc deloc} shows the
evolution of a particular eigenstate with $V_0$. At $V_0 = 0$, the
state is localized due to non-zero $\Delta J$ and exhibits the
characteristic exponential profile. At $V_0 = 22$~kHz, the
diagonal disorder compensates the effects of the off-diagonal
disorder, and the state becomes delocalized. Tuning $V_0$ further
results in the localization of this eigenstate.

\section{Quantum simulation of Holstein polaron model}
\label{s-s-s-holstein polaron}

Rotational Frenkel excitons can be used for quantum simulation of
model Hamiltonians that cannot be engineered with atoms, molecules
or photons as probe particles. Here, we discuss an important
example illustrating the possibility of engineering the Holstein
polaron model with collective excitations of molecules on an
optical lattice.

Polaron is a quasiparticle that describes an electron in a crystal
lattice dressed by lattice phonons. The interaction properties of
polarons are currently researched in an effort to understand the
mechanism of high-${T_{\rm c}}$ superconductivity
\cite{polaron-for-high-Tc} and quantum transport in open quantum
systems \cite{polaron-quantum-transport}. Quasiparticles similar
to polarons can be created by placing an impurity in a Fermi
degenerate gas of ultracold atoms, as demonstrated in several
recent experiments \cite{polaron-exp}. The impurity, produced by
changing the internal state of one of the ultracold atoms, can be
coupled to the Fermi sea via a Feshbach resonance. This gives rise
to Fermi polarons. However, lattice phonons are bosons. Therefore,
a better model of electrons in solid state crystals should be
based on coupling a probe particle to bosons. While this may be
achieved by placing an impurity in a Bose--Einstein condensate of
ultracold atoms, no such experiments have been reported to date.

In a recent study \cite{polaron-paper}, Herrera and Krems showed
that the dipole-dipole interactions between polar molecules on an
optical lattice can be exploited to engineer controllable
couplings between excitons and lattice phonons. In this system,
the phonons are associated with the oscillatory motion of
molecules in the lattice potential
\begin{eqnarray}
\hat V = \sum\limits_n M \omega_0^2 \delta
r_n^2/2 + {\sum\limits_{n, m}}^{'} U_g/|n-m|^3
\label{lattice}
\end{eqnarray}
where $\delta r_n$ is a small deviation from the equilibrium
position of the molecule in site $n$, $M$ is the mass of the
molecules, $\omega_0$ is the trapping frequency, and $U_g = R^3
\langle gg | \hat V_{dd}(R) | gg \rangle$. The first term in
Eq.~(\ref{lattice}) describes uncoupled oscillations of molecules
in their respective lattice sites and depends on the intensity of
the trapping laser that determines the trapping frequency
$\omega_0$. The second term accounts for the collective motion by
coupling molecules in different sites. By introducing the
operators $\hat a_\nu^\dag(k)$ and $\hat a_\nu(k)$ for the phonon
mode $\nu$ with wave vector $k$, we can write the phonon
Hamiltonian as
\begin{eqnarray}
\hat H_{\rm ph} = \sum\limits_{\nu,k} \hbar \omega_{\rm ph}(k)
\hat a_\nu^\dag(k) \hat a_\nu(k),
\end{eqnarray}
where
\begin{eqnarray}
\hbar \omega_{\rm ph}(k) = \omega_0 \sqrt{1 + (12 U_g/a^5 M
\omega_0^2) \sum_{m>0} (1-\cos mk)/m^5}.
\end{eqnarray}
The phonon spectrum is gapped, as $\omega(k \to 0) \to \omega_0$,
and resembles that of optical phonons in solid state crystals.

The exciton--phonon interaction is obtained by expanding the
matrix elements of the dipole-dipole interaction in a Taylor
series to yield:
\begin{equation}
\label{H exc-phonon}
\begin{array}{c}
 \hat H_{\rm exc-ph} =
\sum\limits_{k,\nu; n} g^\nu_F(k) \biggl(\hat a_\nu^\dag(k) + \hat
a_\nu(k)\biggr) \hat P_n^\dag \hat P_n + \\

\\ {\sum\limits_{k,_\nu;
n,m}}^{'} g^\nu_J(k) \biggl(\hat
a_\nu^\dag(k) + \hat a_\nu(k) \biggr) \hat P_n^\dag \hat P_m,\\
\end{array}
\end{equation}
where $g^\nu_F(k) \approx F /a\sqrt{M \omega_0}$, $g^\nu_J(k)
\approx J /a\sqrt{M \omega_0}$, $F = \langle eg | \hat V_{dd}(a)|
eg \rangle - \langle gg | \hat V_{dd}(a)| gg \rangle$, and $J$ is
the excitation transfer matrix element defined in
Eq.~(\ref{Hexc}). The first term in Eq.~(\ref{H exc-phonon})
describes the phonon-modulated transition energies, the second --
the phonon-modulated excitation transfer.

\begin{figure}[ht]
\centering
\includegraphics[scale=0.7]{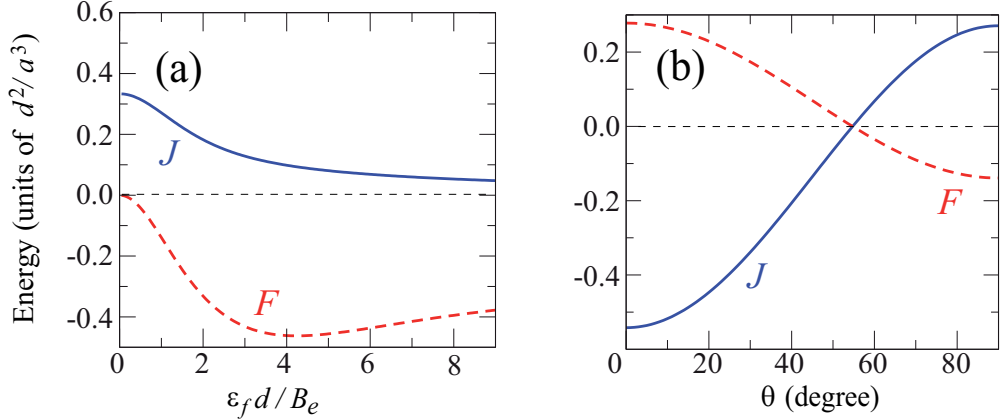}
\caption{$F$ and $J$ as functions of the electric field magnitude
at a fixed angle $\theta = 90^o$ (a) and of the angle $\theta$ at
$d {\cal E}_f = B_e$ (b). Calculations are for a 1D ensemble of
LiCs molecules separated by $a$=400~nm; $d^2/a^3$ = 72 kHz. This
is a modified version of the figure taken from
Ref.\cite{polaron-paper}. \label{f-polaron-FJ}}
\end{figure}

The exciton--phonon coupling constants $g_F$ and $g_J$ depend on
the molecular states $|g \rangle$ and $|e\rangle$, which can be
tuned by an external electric field. The dependence of the
constants $F$ and $J$ on the electric field magnitude ${\cal E}_f$
and direction $\theta$ is shown in Figure \ref{f-polaron-FJ}. In
the limit $g_F \gg g_J$ the Hamiltonian (\ref{H exc-phonon})
describes the Holstein polaron model \cite{Holstein}; this limit
corresponds to large values of ${\cal E}_f$. In the opposite
limit, when $g_F \ll g_J$, it corresponds to the model of
particle--boson coupling by Su, Schrieffer and Heeger \cite{SSH};
this is the limit of small ${\cal E}_f$.

\begin{figure}[ht]
\centering
\includegraphics[scale=0.7]{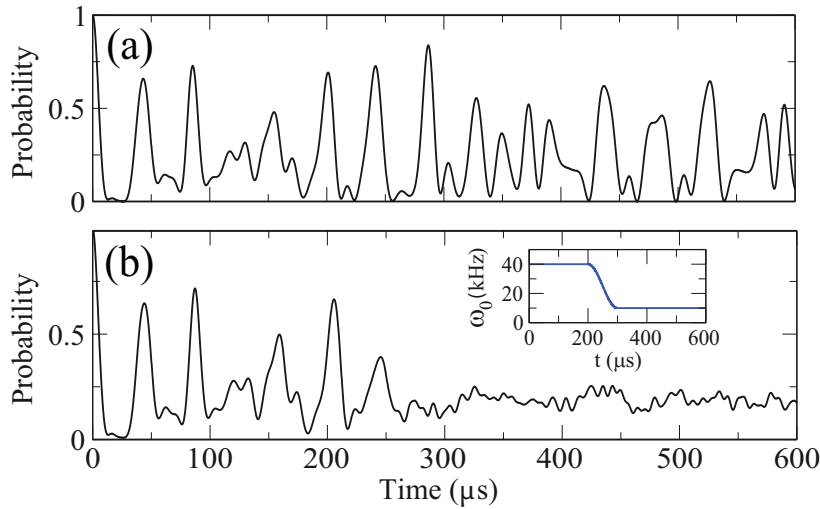}
\caption{Excitation energy transfer in a 1D array of five LiCs
molecules separated by 400 nm in a dc electric field perpendicular
to the array. (a) Evolution of the excitation probability for the
first molecule of the array, when no phonons are present. The
revivals occur due to the finite size of the molecular array and
the reflection of the exciton wavepacket from the edge of the
ensemble. (b) The same as in (a), but with phonons in an optical
lattice with trapping frequency $\nu_0 = \omega_0/2\pi$ varying in
time as indicated in the inset. The field strength is 0.5 kV/cm.
The figure is taken from Ref.\cite{polaron-paper}
\label{f-polaron-dyn}}
\end{figure}

The exciton--phonon coupling for molecules in an optical lattice
can also be controlled by varying the trapping frequency
$\omega_0$, which is proportional to the intensity of the trapping
laser and enters both of the coupling constants as
$1/\sqrt{\omega_0}$. Increasing the exciton--phonon coupling
strength results in an increase of the effective mass of the
polaron. In the limit of strong coupling, polarons are very
massive and localized. The coherent propagation is impeded and
polarons can only propagate by random hopping. By decreasing
$\omega_0$ and therefore increasing the exciton--phonon coupling,
it may be possible to induce self-localization of excitons. The
dynamics of an excitation initially produced on a single molecule
is illustrated in Figure \ref{f-polaron-dyn}. These results
demonstrate that polar molecules trapped on an optical lattice can
be used as a dynamically controllable simulator of open quantum
systems.

\section{Conclusions}

The development of experimental techniques for cooling, trapping
and controlling atoms and molecules has opened exciting
opportunities for new studies of quantum many-body systems.
Ultracold atoms and molecules are extensively researched as
paradigm systems for engineering novel states of quantum matter
and for quantum simulation of model Hamiltonians used in condensed
matter physics. Most of these studies use ultracold atoms or
molecules as probe particles. Here, we discuss the possibility of
using collective excitations in an ensemble of molecules trapped
on an optical lattice as a probe in a quantum simulation
experiment.

We have shown that rotational excitations of polar molecules on an
optical lattice may lead to the formation of collective many-body
excitations -- rotational Frenkel excitons. These excitons have
unique properties that allow tuning the linear and non-linear
exciton interactions by modifying the rotational structure of
ultracold molecules by an external electric field. We suggest that
this can be exploited for the study of new regimes of Frenkel
exciton physics and the dynamics of quantum localization in
disordered systems. We also suggest that rotational Frenkel
excitons can be used for quantum simulation of the Holstein
polaron model. This offers interesting possibilities to study
quantum transport in open quantum systems with controllable
interactions with the environment. In particular, the finite size
and tunable properties of the phonon bath for molecules on an
optical lattice suggests the possibility of exploring the
transition from a non-Markovian to Markovian environment. To
observe rotational excitons, one can measure the populations of
the rotational states at different lattice sites. As described in
Ref.~\cite{DeMille}, this can be achieved by applying a gradient
of an electric field and detecting resonant transitions from
Stark-shifted molecular levels.

\section*{Acknowledgements}
We thank our collaborators, Felipe Herrera, Ping Xiang and
Jes\'{u}s P\'{e}rez-R\'{i}os, who contributed to the original
publications \cite{disordered-paper, polaron-paper,
biexciton-paper, Jesus} forming the basis of this article. Our
work is supported by NSERC of Canada and the Peter Wall Institute
for Advanced Studies at the University of British Columbia.

%===============================================================


\begin{thebibliography}{99}

\bibitem{atom-review}
I.~Bloch, J.~Dalibard, and  W.~Zwerger, Rev. Mod. Phys. 80, 885
(2008)

\bibitem{lincoln-carr-book}
``Understanding Quantum Phase Transitions'', editor: L. D. Carr
(Taylor and Francis, Boca Raton, Fl, 2010)

\bibitem{BEC}
M. H. Anderson, J. R. Ensher, M. R. Matthews, C. E. Wieman, and E.
A. Cornell, Science. 269, 198 (1995)

C. C. Bradley, C. A. Sackett, J. J. Tollett, and R. G. Hulet,
Phys. Rev. Lett. 75, 1687 (1995)

K. B. Davis, M. O. Mewes, M. R. Andrews, N. J. van Druten, D. S.
Durfee, D. M. Kurn, and W. Ketterle, Phys. Rev. Lett. 75, 3969
(1995)

E. A. Cornell and C. E. Wieman, Rev. Mod. Phys. 74, 875 (2002)

W. Ketterle, Rev. Mod. Phys. 74, 1131 (2002)

\bibitem{SF}
M. R. Andrews, D. M. Kurn, H.-J. Miesner, D. S. Durfee, C. G.
Townsend, S. Inouye, and W. Ketterle, Phys. Rev. Lett. 79, 553
(1997)

\bibitem{QMagn}
S. Trotzky, P. Cheinet, S. F\"{o}lling, M. Feld, U. Schnorrberger,
A. M. Rey, A. Polkovnikov, E. A. Demler, M. D. Lukin and I. Bloch,
Science. 319, 295 (2008)

\bibitem{many-body-spin-dyn}
M. Anderlini, P. J. Lee, B. L. Brown, J. Sebby-Strabley, W. D.
Phillips and J. V. Porto, Nature. 448, 452 (2007)

\bibitem{Efimov}
T. Kraemer, M. Mark, P. Waldburger, J. G. Danzl, C. Chin, B.
Engeser, A. D. Lange, K. Pilch, A. Jaakkola, H.-C. N\"{a}gerl and
R. Grimm, Nature. 440, 315 (2006)

C. H. Greene, Physics Today. 63, 40 (2010)

\bibitem{BCS-SF}
J. Kinast, S. L. Hemmer, M. E. Gehm, A. Turlapov, and J. E.
Thomas, Phys. Rev. Lett. 92, 150402 (2004)

C. H. Schunck, Yong-il Shin, A. Schirotzek and W. Ketterle,
Nature. 454, 739 (2008)

\bibitem{BCS-BEC}
C. A. Regal, M. Greiner, and D. S. Jin, Phys. Rev. Lett. 92,
040403 (2004)

M. Greiner, C. A. Regal, and D. S. Jin, Phys. Rev. Lett. 94,
070403 (2005)

C. A. Regal, M. Greiner, S. Giorgini, M. Holland, and D. S. Jin,
Phys. Rev. Lett. 95, 250404 (2005)

\bibitem{trap-atom-in-OL}
I. Bloch, Nature Physics. 1, 23 (2005)

\bibitem{single-atom-detection}
W. S. Bakr, J. I. Gillen, A. Peng, S. F\"{o}lling and M. Greiner,
Nature. 462, 74 (2009)

W. S. Bakr, A. Peng, M. E. Tai, R. Ma, J. Simon, J. Gillen, S.
Foelling, L. Pollet and M. Greiner, Science. 329, 547 (2010)

J. F. Sherson, C. Weitenberg, M. Endres, M. Cheneau, I. Bloch and
S. Kuhr, Nature. 467, 68 (2010)

\bibitem{latt-model-in-sol-st-phys}
M. Lewenstein, A. Sanpera, and V. Ahufinger, ``Ultracold Atoms in
Optical Lattices. Simulating quantum many-body systems'' (Oxford
University Press, USA, 2012)

M. Lewenstein, A. Sanpera, V. Ahufinger, B. Damski, A. Sen and U.
Sen, Advances in Physics. 56, 243 (2007)

\bibitem{MI-SF}
M. Greiner, O. Mandel, T. Esslinger, T. W. H\"{a}nsch and I.
Bloch, Nature. 415, 39 (2002)

\bibitem{high-Tc}
J. G. Bednorz and K. A. M\"{u}ller, Z. Phys. B - Condensed Matter.
64, 189 (1986)

\bibitem{nj-review}
L. D. Carr, D. DeMille, R. V. Krems and J. Ye, New Journal of
Physics. 11, 055049 (2009)

\bibitem{book-chapter}
G. Pupillo, A. Micheli, H.P. B\"{u}chler, and P. Zoller, in ``Cold
Molecules: Theory, Experiment, Applications'', editors: Roman
Krems, Bretislav Friedrich and William C Stwalley (CRC Press,
2006)

\bibitem{N-A-spin-crystal}
A. Micheli, G. K. Brennen and P. Zoller, Nature Physics. 2, 341
(2006)

R. Barnett, D. Petrov, M. Lukin, and E. Demler, Phys. Rev. Lett.
96, 190401 (2006)

G. K. Brennen, A. Micheli and P. Zoller, New J. Phys. 9, 138
(2007)

H. P. B\"{u}uchler, A. Micheli, and P. Zoller, Nature Phys. 3, 726
(2007)

M. L. Wall and L. D. Carr, New J. Phys. 11, 055027 (2009)

C. Trefzger, M. Alloing, C. Menotti, F. Dubin and M. Lewenstein,
New J. Phys. 12, 093008 (2010)

J. P. Kestner, B. Wang, J. D. Sau, and S. D. Sarma, Phys. Rev. B
83, 174409 (2011)

A. V. Gorshkov, S. R. Manmana, G. Chen, J. Ye, E. Demler, M. D.
Lukin, and A. M. Rey, Phys. Rev. Lett. 107, 115301 (2011)

M. Lemeshko, R. V. Krems and H. Weimer,
http://arxiv.org/abs/1203.0010

\bibitem{topological}
M. Lewenstein, Nature physics. 2, 209 (2006)

\bibitem{pmol-in-OL-exp}
T. Volz, N. Syassen, D. M. Bauer, E. Hansis, S. D\"{u}rr and G.
Rempe, Nature Physics. 2, 692 (2006)

A. Chotia, B. Neyenhuis, S. A. Moses, B. Yan, J. P. Covey, M.
Foss-Feig, A. M. Rey, D. S. Jin and J. Ye, Phys. Rev. Lett. 108,
080405 (2012)

\bibitem{soliton}
S. Burger, K. Bongs, S. Dettmer, W. Ertmer, K. Sengstock, A.
Sanpera, G. V. Shlyapnikov, and M. Lewenstein, Phys. Rev. Lett.
83, 5198 (1999)

U. Al Khawaja, H. T. C. Stoof, R. G. Hulet, K. E. Strecker, and G.
B. Partridge, Phys. Rev. Lett. 89, 200404 (2002)

K. E. Strecker, G. B. Partridge, A. G. Truscott, and R. G. Hulet,
Nature. 417, 150 (2002)

R. Balakrishnan, I. I. Satija and C. W. Clark, Phys. Rev. Lett.
103, 230403 (2009)

\bibitem{roton}
D. H. J. O'Dell, S. Giovanazzi, and G. Kurizki, Phys. Rev. Lett.
90, 110402 (2003)

S. Sinha and G. V. Shlyapnikov, Phys. Rev. Lett. 94, 150401 (2005)

S. C. Cormack, D. Schumayer, and D. A. W. Hutchinson, Phys. Rev.
Lett. 107, 140401 (2011)

\bibitem{vortex}
K. W. Madison, F. Chevy, W. Wohlleben, and J. Dalibard, Phys. Rev.
Lett. 84, 806 (2000)

M. W. Zwierlein, J. R. Abo-Shaeer, A. Schirotzek, C. H. Schunck
and W. Ketterle, Nature. 435, 1047 (2005)

A. Nunnenkamp, A. M. Rey and K. Burnett, Proc. R. Soc. A. 466,
1247 (2010)

\bibitem{magnon}
H. J. Lewandowski, D. M. Harber, D. L. Whitaker and E. A. Cornell,
Phys. Rev. Lett. 88, 070403 (2002)

J. M. McGuirk J M, H. J. Lewandowski, D. M. Harber, T. Nikuni, J.
E. Williams and E. A. Cornell, Phys. Rev. Lett. 89, 090402 (2002)

\bibitem{polaron-exp}
A. Schirotzek, C.-H. Wu, A. Sommer, and M. W. Zwierlein, Phys.
Rev. Lett. 102, 230402 (2009)

S. Nascimb\`{e}ne, N. Navon, K. J. Jiang, L. Tarruell, M.
Teichmann, J. McKeever, F. Chevy, and C. Salomon, Phys. Rev. Lett.
103, 170402 (2009)

C. Kohstall, M. Zaccanti, M. Jag, A. Trenkwalder, P. Massignan, G.
M. Bruun, F. Schreck, and R. Grimm,
http://arxiv.org/pdf/1112.0020.pdf

\bibitem{whaley-pccp}
R. E. Zillich and K. B. Whaley, Phys. Chem. Chem. Phys. 13, 18835
(2011)

\bibitem{carr-molecular-hubbard}
M. L. Wall and L. D. Carr, Phys. Rev. A. 82, 013611 (2010)

\bibitem{gorshkov-superfluidity}
K. A. Kuns, A. M. Rey and A. V. Gorshkov, Phys. Rev. A. 84, 063639
(2011)

\bibitem{QS-reviews}
J.~I.~Cirac and P.~Zoller, Nature Physics. 8, 264 (2012)

J. Buluta and F. Nori, Science. 326, 108 (2009)

\bibitem{feynman}
R. P. Feynman,International Journal of Theoretical Physics. 21,
467, 1982

\bibitem{Manin}
Yu. I. Manin, ``Computable and non-computable'' (Moscow,
"Sovetskoe radio", 1980, p. 15), in Russian.

\bibitem{QS-atoms}
I. Bloch, J. Dalibard and S. Nascimb\`{e}ne, Nature Physics. 8,
267 (2012)

\bibitem{QS-ions}
R. Blatt and C. F. Roos, Nature Physics. 8, 277 (2012)

\bibitem{QS-photons}
A. Aspuru-Guzik and P. Walther, Nature Physics. 8, 285 (2012)

\bibitem{QS-Rydberg}
H. Weimer, M. M{\"u}ller, I. Lesanovsky, P. Zoller and H. P.
B{\"u}chler, Nature Physics. 6, 382 (2010)

\bibitem{QS-circuit}
A. A. Houck, H. E. T{\"u}reci and J. Koch, Nature Physics. 8, 292
(2012)

\bibitem{QS-NV-centers}
R. Hanson and D. D. Awschalom, Nature. 453, 1043 (2008)

\bibitem{review-simulation-with-atoms}
J. Simon, W. S. Bakr, R. Ma, M. E. Tai, P. M. Preiss and M.
Greiner, Nature. 472, 307 (2011)

\bibitem{optical-lattice}
M. Greiner and S. F\"{o}lling, Nature. 453, 736 (2008)

\bibitem{single-site-detection}
C. Weitenberg, M. Endres, J. F. Sherson, M. Cheneau, P. Schau\ss,
T. Fukuhara, I. Bloch, and S. Kuhr, Nature. 471, 319 (2011)

\bibitem{experiments}
F. H. L. Essler, H. Frahm, F. G\"{o}hmann, A. Kl\"{u}mper and V.
E. Korepin, ``The One-Dimensional Hubbard Model'' (Cambridge
University Press, Cambridge, 2005)

D. Jaksch, C. Bruder, J. I. Cirac, C. W. Gardiner and P. Zoller,
Phys. Rev. Lett. 81, 3108 (1998)

M. K\"{o}hl, H. Moritz, T. St\"{o}ferle, K. G\"{u}nter, T.
Esslinger, Phys. Rev. Lett. 94, 080403 (2005)

\bibitem{HubMod-for-highTc}
P. W. Anderson, P. A. Lee, M. Randeria, T. M. Rice, N. Trivedi and
F. C. Zhang, J. Phys.: Condens. Matter. 16, R755 (2004)

\bibitem{HubMod-for-magnetism}
M. Lewenstein and A. Sanpera, Science. 319, 292 (2008)

\bibitem{superfluidity}
R. Micnas and B. Tobijaszewska, J. Phys.: Condens. Matter. 14,
9631 (2002)

\bibitem{mott}
H. Heiselberg, Phys. Rev. A. 73, 013628 (2006)

\bibitem{99-percent}
W. S. Bakr, P. M. Preiss, M. Eric Tai, R. Ma, J. Simon and M.
Greiner, Nature. 480, 5003 (2011)

\bibitem{pmol-for-long-range}
K. G\'{o}ral, L. Santos and M. Lewenstein, Phys. Rev. Lett. 88,
170406 (2002)

\bibitem{pmol-review}
D. S. Jin and J. Ye, Physics Today. 64, 27 (2011)

\bibitem{tonks}
L. Tonks, Phys. Rev. 50, 955 (1936)

\bibitem{girardeau}
M. Girardeau, J. Math. Phys. 1, 516, (1960)

\bibitem{1D-boson-review}
M. A. Cazalilla, R. Citro, T. Giamarchi, E. Orignac and M. Rigol,
Rev. Mod. Phys. 83, 1405 (2011)

\bibitem{xxz}
R. Orbach, Phys. Rev. 112, 309 (1958)

\bibitem{kaplan}
I. G. Kaplan, Theor. Math. Phys. 27, 466 (1976)

\bibitem{agranovich-book}
V. M. Agranovich, ``Excitations in Organic Solids'' (Oxford
University Press, Oxford, 2008)

\bibitem{jordan-wigner}
P. Jordan and E. Wigner, Z. Phys. 47, 631 (1928)

\bibitem{holstein-primakoff}
T. Holstein and H. Primakoff, Phys. Rev. 58, 1098 (1940)

\bibitem{chestnut-suna}
D. B. Chesnut and A. Suna, J. Chem. Phys. 39, 146 (1963)

\bibitem{agranovich-toshich}
V. M. Agranovich and B. S. Toshich, Zh. Eksp. Teor. Fiz. 53, 149
(1967)

\bibitem{ML}
M. Litinskaya, Phys. Rev. B. 77, 155325 (2008)

\bibitem{disordered-paper}
F. Herrera, M. Litinskaya and R. V. Krems, Phys. Rev. A. 82,
033428 (2010)

\bibitem{Ya.-I.-Frenkel}
J. Frenkel, Phys. Rev. 37, 17 (1931)

\bibitem{beyond-Heit-Lon}
V. M. Agranovich and D. M. Basko, J. Chem. Phys. 112, 8156 (2000)

\bibitem{negative-refraction}
V. M. Agranovich and Yu. N. Gartstein, ``Spatial dispersion and
negative refraction of light'', UFN (Physics: Uspekhi). 176, 1051
(2006)

\bibitem{polaron-paper}
F. Herrera and R. V. Krems, Phys. Rev. A. 84, 051401(R) (2011)

\bibitem{biexciton-paper}
P. Xiang, M. Litinskaya and R. V. Krems, Phys. Rev. A 85,
061401(R) (2012)

\bibitem{vektaris}
G. Vektaris, J. Chem. Phys. 101, 3031 (1994)

\bibitem{rabl-zoller-paper}
P. Rabl and P. Zoller, Phys. Rev. A. 76, 042308 (2007)

\bibitem{dyson}
F. J. Dyson, Phys. Rev. 102, 1217 (1956)

\bibitem{Anderson-loc-review}
See for a review: A. Lagendijk, B. van Tiggelen and D. S. Wiersma,
Physics Today. 62, 24 (2009)

\bibitem{disorder-light}
D. S. Wiersma, P. Bartolini, A. Lagendijk and R. Righini, Nature.
390, 671 (1997)

\bibitem{disorder-matter-waves}
J. Billy, V. Josse, Z. Zuo, A. Bernard, B. Hambrecht, P. Lugan, D.
Cl\'{e}ment, L. Sanchez-Palencia, P. Bouyer and A. Aspect, Nature.
453, 891-894 (2008)

\bibitem{isulator-conductor-in-solid}
J. Chab\'{e}, G. Lemari\'{e}, B. Gr\'{e}maud, D. Delande, P.
Szriftgiser and J. C. Garreau, Phys. Rev. Lett. 101, 255702 (2008)

\bibitem{glass}
K. Binder and A. P. Young, Rev. Mod. Phys. 58, 801 (1986)

\bibitem{supercond-insulator}
N. Mason and A. Kapitulnik, Phys. Rev. Lett. 82, 5341 (1999)

N. Markovic, C. Christiansen, A. M. Mack, W. H. Huber and A. M.
Goldman, arXiv:cond-mat/9904168v1.

\bibitem{disorder-in-quasicrystals}
L. Levi, M. Rechtsman, B. Freedman, T. Schwartz, O. Manela and M.
Segev, Science. 332, 1541 (2011)

\bibitem{zajman}
J. M. Ziman, ``Models of Disorder: The Theoretical Physics of
Homogeneously Disordered Systems'' (Cambridge University Press,
1979)

\bibitem{IMLifshitz}
I. M. Lifshitz, Zh. Eksp. Teor. Fiz. 18, 293 (1948)

I. M. Lifshitz and V. I. Peresada, Uchenye zapiski Kharkovskogo
Universiteta Vol. {\bf 64}; Trudy Fiz. otd. fiz.-mat. f-ta. 6, 37
(1955), in Russian

\bibitem{dubovsky}
O. A. Dubovsky and Yu. V. Konobeev, Fiz. Tverd. Tela (Leningrad).
7, 946 (1965)

\bibitem{dubovsky-2}
O. A. Dubovsky and Yu. V. Konobeev, Fiz. Tverd. Tela (Leningrad).
6, 946 (1965)

\bibitem{la-rocca}
H. Zoubi and G. C. La Rocca, Phys. Rev. B. 72, 125306 (2005)

\bibitem{Ioffe-Regel}
A. F. Ioffe and A. R. Regel, Prog. Semicond. 4, 237 (1960)

\bibitem{weakloc}
G. Bergmann, Phys. Rep. 107, 1 (1984)

D. E. Khmelnitskii, Physica B (Amsterdam). 126, 235 (1984)

\bibitem{1D-loc}
H. Fukuyama and S. Hikami, ``Anderson Localization'' (Springer,
Berlin, 1982)

\bibitem{deloc-discr}
J. C. Flores, J. Phys. Cond. Matt. 1, 8471 (1989)

A. Bovier, J. Phys. A. 25, 1021 (1992)

\bibitem{deloc-contin}
F. M. Izrailev and A. A. Krokhin, Phys. Rev. Lett. 82, 4062 (1999)

\bibitem{tesseri}
L. Tessieri and F. M. Izrailev, Physica E. 9, 405 (2001)

\bibitem{polaron-for-high-Tc}
A. S. Alexandrov and N. F. Mott, ``Polarons and Bipolarons''
(World Scientific Publishing, Singapore, 1995)

\bibitem{polaron-quantum-transport}
F. Caruso, A. W. Chin, A. Datta, S. F. Huelga and M. B. Plenio, J.
Chem. Phys. 131, 105106 (2009)

P. Rebentrost, M. Mohseni, I. Kassal, S. Lloyd and A.
Aspuru-Guzik, New J. Phys. 11, 033003 (2009)

D. Segal and D. R. Reichman, Phys. Rev. A. 76, 012109 (2007)

\bibitem{Holstein}
T. Holstein, Ann. Phys. (NY). 8, 325 (1959)

\bibitem{SSH}
W. P. Su, J. R. Schrieffer and A. J. Heeger, Phys. Rev. Lett. 42,
1698 (1979)

\bibitem{DeMille}
D. DeMille, Phys. Rev. Lett. 88, 067901 (2002)

\bibitem{Jesus}
J. P\'{e}rez-R\'{i}os, F. Herrera and R. V. Krems, New J. Phys.
12, 103007 (2010)

\end{thebibliography}
\end{document}